\documentclass{emulateapj}
\usepackage{epsfig}
\usepackage{amsmath}
\usepackage{amssymb}
\usepackage{natbib}
\usepackage{setspace}
\usepackage{graphicx}
\begin{document}
\title{Short GMC lifetimes: an observational estimate with the PdBI Arcsecond Whirlpool
  Survey (PAWS)}
\author{Sharon E. Meidt\altaffilmark{1}, Annie Hughes\altaffilmark{1}, Clare L. Dobbs\altaffilmark{2}, J\'er\^ome Pety\altaffilmark{3}$^,$\altaffilmark{4}, Todd A. Thompson\altaffilmark{8}$^,$\altaffilmark{9}, Santiago Garc\'{i}a-Burillo\altaffilmark{5}, Adam K. Leroy\altaffilmark{6}, Eva Schinnerer\altaffilmark{1}, Dario Colombo\altaffilmark{1}, Miguel Querejeta\altaffilmark{1}, Carsten Kramer\altaffilmark{7}, Karl F. Schuster\altaffilmark{4}, Ga\"elle Dumas\altaffilmark{4}}

\altaffiltext{1}{Max-Planck-Institut f\"ur Astronomie / K\"{o}nigstuhl 17 D-69117 Heidelberg, Germany}
\altaffiltext{2}{School of Physics and Astronomy, University of Exeter, Stocker Road, Exeter EX4 4QL, UK}
\altaffiltext{3}{Institut de Radioastronomie Millim\'etrique, 300 Rue de la Piscine, F-38406 Saint Martin d'H\`eres, France}
\altaffiltext{4}{Observatoire de Paris, 61 Avenue de l'Observatoire, F-75014 Paris, France.}
\altaffiltext{5}{Observatorio Astron\'{o}mico Nacional - OAN, Observatorio de Madrid Alfonso XII, 3, 28014 - Madrid, Spain}
\altaffiltext{6}{National Radio Astronomy Observatory, 520 Edgemont Road, Charlottesville, VA 22903, USA}
\altaffiltext{7}{Instituto Radioastronom\'{i}a Milim\'{e}trica, Av. Divina Pastora 7, Nucleo Central, 18012 Granada, Spain}
\altaffiltext{8}{Department of Astronomy, The Ohio State University, 140 W. 18th Ave., Columbus, OH 43210, USA} 
\altaffiltext{9}{Center for Cosmology and AstroParticle Physics, The Ohio State University, 191 W. Woodruff Ave., Columbus, OH 43210, USA}

\begin{abstract}
We describe and execute a novel approach to observationally estimate the lifetimes of giant molecular clouds (GMCs).  We focus on the cloud population in the zone between the two main spiral arms in M51, i.e. the inter-arm region, where cloud destruction via shear and star formation feedback dominates over cloud formation processes.   By monitoring the change in GMC number densities and ensemble properties from one side of the inter-arm to the other, we estimate the cloud lifetime as a fraction of the inter-arm travel time.  
We find that cloud lifetimes in M51's inter-arm are finite and short, i.e. 20 to 30 Myr.  Such short lifetimes suggest that cloud evolution is influenced by environment, in which processes are sufficient to disrupt GMCs after a few free-fall times.  Over most of the region under investigation shear appears to regulate cloud lifetimes.  
As the shear timescale increases with galactocentric radius, we expect cloud destruction to switch primarily to star formation feedback at larger radii.  We identify a transition from shear-dominated to star formation feedback-dominated cloud disruption through a change in the behavior of the inter-arm GMC number density.  The signature suggests that shear is more efficient at completely dispersing clouds, whereas star formation feedback tends to transform the cloud population, e.g. by fragmenting high mass clouds into lower mass pieces. 
Compared to the characteristic timescale for molecular hydrogen in M51, our short cloud lifetime measurements suggest that gas can remain molecular while clouds disperse and reassemble.  We propose that galaxy dynamics regulates the cycling of molecular material from diffuse to bound -- and ultimately star-forming -- objects, and hence contributes to long observed molecular gas depletion times in normal disk galaxies.  We also speculate that, in more extreme environments such as elliptical galaxies and concentrated galaxy centers, star formation can be suppressed when the shear timescale becomes so short that some clouds never have the opportunity to collapse and form stars.  
\end{abstract}

\maketitle
\date{\today}
%%%%%%%%%%%%%%%%%%%%%%%%%%%%%%\doublespacing

\section{Introduction}
The lifetimes of giant molecular clouds (GMCs) set a natural limit to the timescale over which gas is converted into stars 
 and thus potentially impose powerful constraints on simulations of galaxy formation and evolution.   
Yet very little is directly observationally known about cloud longevity.  
Recently, a wealth of observations at high spatial resolution and sensitivity in nearby galaxies have stimulated renewed debate on the issue of cloud lifetimes, their relation to the molecular content of galaxies and the process of star formation at all spatial scales.    

The very fact that we observe clouds in spiral galaxies populating the area between spiral arms (in the so-called inter-arm region), has been taken as evidence that they are long-lived, surviving at least as long as the $\sim$100 Myr required to pass from one spiral to the next (\citealt{sss79}; \citealt{scovilleWilson}; \citealt{koda09}).  Cloud longevity -- and the corollary notion that GMCs are supported against gravitational collapse by turbulence and/or magnetic fields-- is thought to explain why molecular gas depletion times greatly exceed the free-fall time of an individual GMC (e.g. \citealt{fleck}; \citealt{shu87}; \citealt{kmm06}).  

In contrast, observational reconstruction of cloud life-cycles in the LMC indicates a much shorter lifetime of $\sim30$\,Myr \citep{kawamura}, in good agreement with earlier studies that applied a
similar technique to Milky Way GMCs (e.g. \citealt{bash}, \citealt{Leisawitz}).  Using the same approach, \cite{miura} also
obtained a typical GMC lifetime of 20 to 40 Myr in M33. 
Such short lifetimes are consistent with recent models of GMC formation and evolution that predict lifetimes nearer 10-20 Myr (e.g. \citealt{dbp11}; \citealt{dpb12}; \citealt{dobbsP13}).  In simulations with realistic distributions of gas and stars, ISM heating/cooling and star formation, the combination of shear and star formation feedback lead to fairly rapid cloud dispersal (\citealt{dpb12}; \citealt{dobbsP13}).  In support of this picture, two recent studies of the cloud population in M51 conclude that clouds are far from  `standard' and do not obey the scaling relations normally taken to imply that they are virialized objects (\citealt{hughesPDFs}; \citealt{colombo2014a}).  This suggests that clouds are not necessarily stable, long-lived entities.  

Whether clouds are long- or short-lived has strong implications for the efficiency of star formation observed across a range of spatial scales (\citealt{kt07}; \citealt{feldmann}), impacting our view of the balance between cloud-scale physical processes thought to regulate star formation, e.g. turbulent driving, magnetic fields and feedback in the form of mechanical and thermal energy (\citealt{VS03}; \citealt{km05}; \citealt{elm07}; Kim et al. 2011; \citealt{VS05}; \citealt{pricebate}).  
In order to distinguish between the short-lived/dynamic versus long-lived/quasi-static views of GMC evolution, a greater number of direct observational estimates for GMCs lifetimes in systems beyond the Local Group is urgently required. To date, the high resolution data required to establish an empirical picture of the GMC lifecycle has only been available for the Milky Way, the Magellanic Clouds and M33, and hence the diversity of galactic environments for which there are empirical estimates of GMC lifetimes is very limited.

In this paper, we describe a technique for estimating GMC
lifetimes in star-forming disk galaxies that is independent of their association
with young stellar phenomena. We exploit the large number statistics
provided by the PdBI Arcsecond Whirlpool Survey (PAWS; \citealt{schinner2013}, \citealt{petyPAWS}) to monitor changes in the GMC population
across the inner disk of M51, which we then interpret using our
detailed knowledge of gas dynamics and high mass star formation across
the PAWS field. The premise is a simple one: a short (i.e. less than an orbital period) cloud lifetime should manifest itself as a decrease in the number of clouds from one side of the inter-arm to the other.  The ratio of the initial number of clouds to the number of clouds `lost' during inter-arm passage constrains the ratio of the cloud lifetime to the time to traverse the inter-arm. 
Since the travel time in the inter-arm increases with galactocentric radius the fraction of `lost clouds'  may exhibit a radial dependence: fewer clouds are expected to survive over the full passage through the inter-arm at larger galactocentric radius. The signature may also depend on the dominant cloud disruption mechanism (shear, star formation feedback) within a given zone of the galactic disk.  

This paper is organized as follows. In $\S$
\ref{sec:framework}, we present our method to estimate cloud
lifetimes. As we are primarily interested in cloud destruction processes and the limits they impose on cloud lifetimes, we focus on the cloud population in the inter-arm region of disk galaxies, i.e. the zone in between the  main spiral arms.  Compared to the high density
arm environment, fewer clouds are expected to form in the inter-arm
and, moreover, the dynamics in this region of the disk are easier to
characterize.  In $\S$ \ref{sec:strengths}, we
summarize the strengths and weaknesses of our method compared to other
approaches for estimating cloud lifetimes and discuss its general applicability.  In $\S$ \ref{sec:appM51}, we apply our framework to M51,
where we use the pattern of star formation and shear in M51's inter-arm
zone to separately study the mechanisms that act to limit the
lifetimes of clouds. 
We present our measurements of cloud lifetimes in
M51's inner disk in $\S$ \ref{sec:measurements}. 
 We discuss and interpret radial trends in cloud lifetimes and cloud destruction
mechanisms in $\S$ \ref{sec:interpretation}.   
Finally, in $\S$ \ref{sec:discussion} we relate our findings to previous cloud lifetime estimates, emphasizing their impact on our view of cloud-scale star formation.  

\section{The measurement framework}
\label{sec:framework}
In this section we outline our method to estimate cloud lifetimes, 
presenting first a heuristic and then a more quantitative description
of our approach. Since we aim to provide an analysis framework
for application to real data, we briefly discuss the connection
between the idealized clouds in our model and GMCs as identified in CO
observations.  We refer to the PAWS CO(1-0) survey to illustrate these
general considerations, which are common to all cloud-scale
extragalactic CO surveys to which our method can be applied. Later in $\S$ \ref{sec:appM51} we apply our
technique to the inter-arm cloud population in M51 to derive an
estimate of the characteristic GMC lifetime.  

\subsection{Cloud formation and destruction in disk galaxies}

\label{sec:sourceSinks}

The key element for measuring cloud lifetimes with our method is a survey of molecular clouds in the inter-arm zone between spiral arms.  This allows us to track azimuthal variations in the inter-arm GMC number density, which we hypothesize reflect the balance between cloud destruction and formation processes
within the inter-arm zone.  
This is motivated by our analysis of GMCs identified in the PAWS survey of CO(1-0) emission in the central 9 kpc of M51 
\citep{colombo2014a} 
which showed that the   
number of inter-arm clouds in the inner disk of M51 decreases from the
downstream of one spiral arm (henceforth ``zone I'', see Figure~1), to
the upstream region of the other arm (``zone II'').  The cloud mass spectrum also changes in shape across the inter-arm, consistent with the idea that the cloud population evolves across the inter-arm due to disruptive processes such as shear and feedback \citep{colombo2014a}.  

\begin{figure}[t]
\begin{center}
\includegraphics[width=1.1\linewidth]{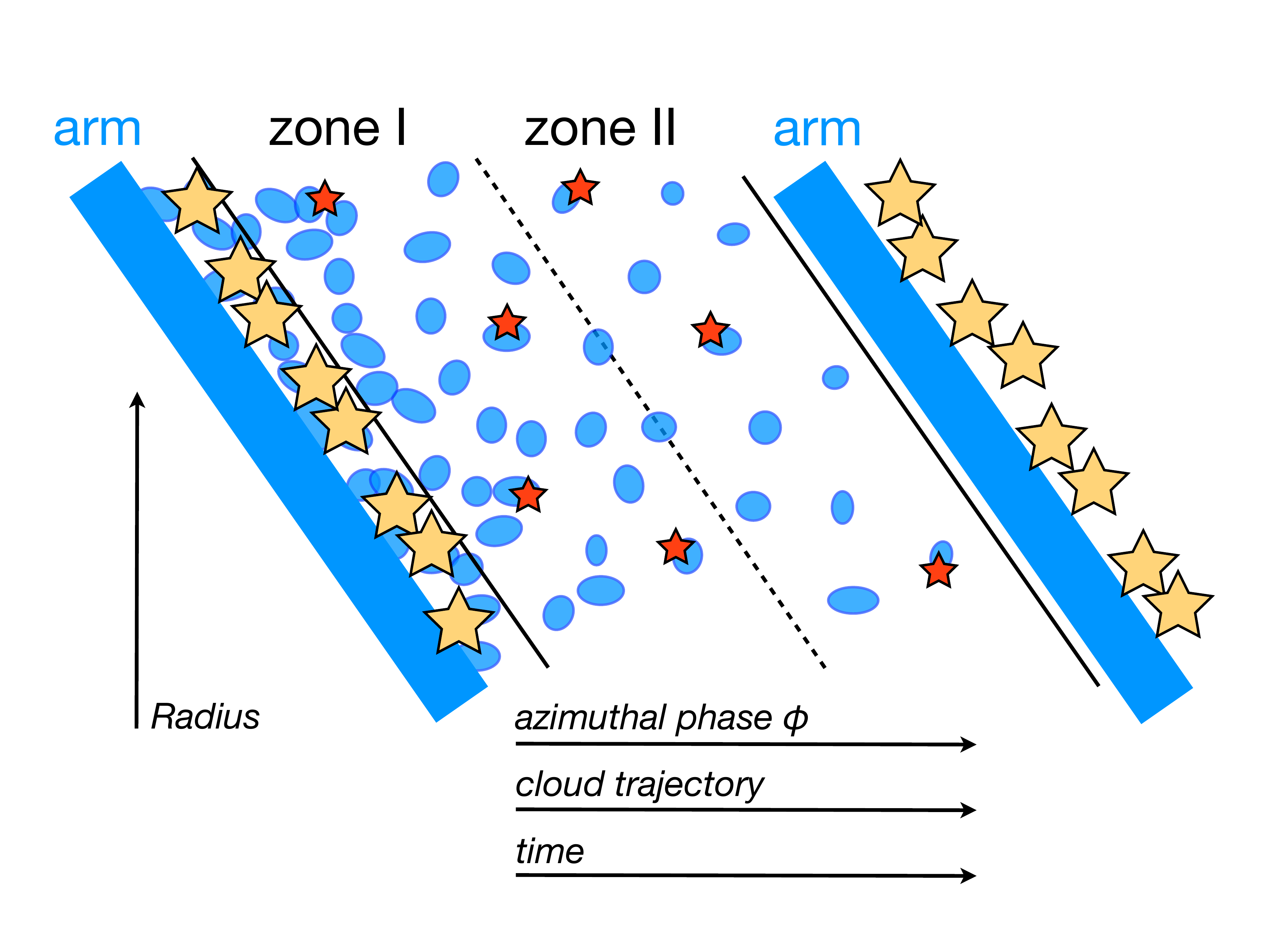}
\caption{Polar coordinate representation of the inter-arm cloud population (blue ellipses) between two spiral arms (blue diagonal lines).  The two solid black lines mark the boundaries of the inter-arm environment.  The black dashed line shows the midpoint that divides the inter-arm into zone I and zone II.  The distribution of star formation events (stars) qualitatively follows the observed pattern of star formation in M51 (see Figure \ref{fig:map}): orange stars represent star formation located within the spiral arm environment, while red stars indicate inter-arm star formation.  In this representation, the trajectory of clouds through the inter-arm is in the horizontal direction (as indicated by the arrows). }
\label{fig:sketch}
\end{center}
\end{figure}

Hereafter, we adopt the following simple picture for the passage of 
clouds through the inter-arm (shown in Figure \ref{fig:sketch}): after leaving the downstream of
one spiral arm, clouds enter zone I and proceed to zone II, which is
located upstream of the next spiral arm.  Under the assumption that
clouds follow roughly circular paths through the inter-arm (see
$\S$ \ref{sec:cloudtrajectories}) the current azimuthal
position of a cloud provides a measure of the time that has passed since it
entered the inter-arm.  Using the galaxy's orbital period as a fiducial
clock, we can thus connect evolution in the number of clouds from zone~I to zone~II with an estimate of the characteristic GMC
lifetime.  

As clouds pass from zone I to zone II in the inter-arm, we expect them to be susceptible to two primary destructive mechanisms, shear and star formation feedback.  
Shear acts on all clouds in
galactic disks undergoing differential rotation, and thus may be
considered to set the `natural' cloud lifetime in the absence of other
destructive processes.  Star formation feedback should also contribute
to cloud disruption, although the timescale associated with
destruction via feedback may depend on the specific energetics of
the star formation event and a cloud's proximity to the star-formation
activity. 

These two processes may lead to different signatures in the variation of GMC
number density with azimuth (although our estimate of the GMC
lifetime does not rely on it).  
Shear, for example, may lead to mass loss and/or complete cloud dispersal \citep{dobbsP13}.  In the presence of shear, our basic
expectation is that the number of clouds and their total combined mass
smoothly decreases from zone~I to zone~II, as material in clouds is
returned to the ISM.  By contrast, feedback, which can disrupt clouds as well as sweep up and compress interstellar gas 
(e.g. \citealt{dawson}), is expected to be spatially localized and
coincident with tracers of high mass star formation.  
In this case, mass loss may be less gentle and potentially involve cloud splitting \citep{dbp11}, thus forming new clouds.  At the same time, feedback acting externally can produce converging flows, possibly merging pre-existing cloud fragments \citep{dpb12} and increasing cloud masses. 
As a result, we expect feedback-dominated zones to 
exhibit a transformation of the cloud population (i.e. an exchange of
mass between high- and low-mass clouds) near sites of star-formation,
rather than a steady decline in both the number and combined mass of
clouds. 

In either case, these considerations suggest that some modes
of cloud destruction can be accompanied by an {\it increase} in clouds
(usually an increase in the number of low-mass clouds), and not solely
a decrease in the total number of clouds.
In our model of GMC evolution in the inter-arm region, we therefore
consider both sinks and sources, i.e. cloud destruction and cloud
formation. \\

\subsection{Measuring the cloud lifetime}

We now present a more formal description of our method for estimating the characteristic GMC lifetime.  Although we cannot directly follow the evolution of clouds with only a single observational snapshot,  
we can adopt an Eulerian representation of the gas flow given a sufficient number of clouds in our sample.  This allows us to replace measurements of clouds throughout their lifetimes (Lagrangian representation) by statistical cloud measurements as a function of position in a spiral galaxy (in our case M51). 
In this way, for a cloud population of initial size $N_0$ at time $t_0$ that undergoes subsequent evolution, we can estimate the cloud lifetime $\tau$ using a measurement of the cloud population at some later time $t_1$.   Here we assume that a single lifetime $\tau$ (the statistical average in the population) applies to all clouds, rather than a distribution of cloud lifetimes.  In this case, if the population is reduced to $N_{1}$ by time $t_1$, then the implied loss rate of $-(N_0-N_{1})/(t_1-t_0)$ leaves no clouds  
left after time $\tau$, i.e.
\begin{equation}
N_{0}-\frac{N_0-N_{1}}{t_1-t_0}\tau=0. \label{eq:start}
\end{equation}

Although we assume that the population overall suffers losses, we let the rate $-(N_0-N_{1})/(t_1-t_0)$ represent the effective rate of change in the population in the presence of sources and sinks, assuming that the loss and gain (or growth) rates are independent of time, i.e.
\begin{equation}
\frac{N_0-N_{1}}{t_1-t_0}=(loss-gain)=N_0 \left(\frac{1}{\tau_{true}}-\frac{1}{\tau_{grow}}\right) \label{eq:startb}
\end{equation}
Here we specifically let $\tau_{grow}$ represent the time to increase the population by $N_0$ at the current growth rate and take $\tau_{true}$ to be the time it would take to reduce the initial population at the given loss rate to zero, i.e. the true cloud lifetime.   

Eqs. (\ref{eq:start}) and (\ref{eq:startb}) together imply
\begin{equation}
\frac{1}{\tau}=\frac{1}{\tau_{true}}-\frac{1}{\tau_{grow}} \label{eq:timescales}
\end{equation} 
and we see that when $\tau_{grow}$$>>$$\tau_{true}$, the cloud lifetime can be approximated by $\tau$. 

In principle, measurements $N_1$ and $N_0$ can be made in any two adjacent zones between which there is a well-defined travel time $t_1-t_0$.  Since we use clouds sampled at different spatial locations along a common trajectory as proxies for the evolution of a single cloud, the travel time between two zones separated by angular distance $\Delta\phi$ represents the Lagrangian-equivalent time for a cloud to traverse that distance, from one zone to the next.  

To estimate the travel time requires knowledge of the angular velocity $V_\phi$.  In the inter-arm region, where non-circular motions are negligible, clouds follow a circular path and $V_\phi$ can be approximated by the circular velocity in the disk (see also $\S$ \ref{sec:traveltime} for detailed discussion). Positioning the measurement zones in this region, to avoid the spiral arms, leads to more reliable estimation of $t_{trav}$ than when zones overlap with spiral arms, where non-circular motions are present and must be taken into account. Focussing on the inter-arm also offers a more direct measure of $t_{true}$, since clouds here are expected to Äundergo relatively less formation compared to, i.e. in the spiral arms.  

Given large enough numbers of clouds, in principle more than two zones in the inter-arm can be considered, providing more than one independent measure of the lifetime.  But to maximize the number of clouds in each zone, we recommend a single lifetime measurement (per inter-arm) using two zones that together span the entire width of the inter-arm.  In this case, eq. (\ref{eq:start}) simplifies to 
\begin{equation}
\tau=\frac{t_{travel}}{2}\frac{N_{I}}{N_{I}-N_{II}}.\label{eq:num}
\end{equation}
Here, the cloud population spanning some (azimuthal) area $\Delta\phi$ at a snapshot in time is divided in half, into two populations, $N_{0}$=$N_{I}$ and $N_{1}$=$N_{II}$.  Measurements in the two halves (zones) are thus separated in time by the equivalent time in a Lagrangian representation for a test cloud to cross half the distance $\Delta\phi$ (i.e. $t_{travel}$/2 where $t_{travel}$ is the time to cross the full distance).  

In terms of the lost fraction $F_{lost}=(N_{I}-N_{II})/N_{I}$ 
eq. (\ref{eq:num}) becomes
\begin{equation}
\tau=\frac{t_{travel}}{2}\frac{1}{F_{lost}}\label{eq:numlost}, 
\end{equation}
and we expect short cloud lifetimes to result in higher $F_{lost}$, or fewer clouds in zone II compared to zone I.  

In our recommended application of eq. (\ref{eq:num}), measurements of the cloud number density in a spiral galaxy occur in the two halves of the inter-arm, zone I or zone II, so that  $t_{travel}$ is the  length of time to cross the distance $\Delta\phi$ between spiral arms at their present location.\footnote{Although eqs. (\ref{eq:num}) and (\ref{eq:numlost}) can be applied, in principle, to galaxies without strong spiral patterns (or even outside the inter-arm) provided an appropriate $t_{travel}$ can be defined, in practice the estimated lifetime is highly uncertain: the cloud populations of flocculent galaxies or those with weak, multi-armed spirals tend to be sparser than in galaxies with strong, well-defined spiral arms, which provide a source of new clouds.  Without large numbers of clouds, the method cannot be reliably applied.  }  The only requirement is that the azimuthal distance $\Delta\phi$ should be large enough that the corresponding travel time spans enough of a cloud's evolution to be sensitive to factors that limit its lifetime.  Note that the travel time defined in this way is shorter than the full length of time for a cloud to pass from one spiral arm to the next, $t_{sp}=2\pi/m(\Omega-\Omega_p)$ where $\Omega_p$ is the so-called pattern speed of the spiral and $m$ is the number of arms.  

In practice, relating evolution in the number of clouds from one side of the inter-arm ($N_{I}$) to the other ($N_{II}$) with cloud lifetimes using eq. (\ref{eq:num}) requires three key components: (1) a catalog of cloud
positions and properties that is complete to a well-determined
sensitivity limit; (2) an accurate measure of the rotation curve (to estimate the inter-arm travel time as a function of galactocentric radius); and (3) an estimate of radial streaming motions (to place a bound on the radial excursion of clouds in their orbits as they travel through the inter-arm).  Each of these should be readily available for most (nearby) galaxies with existing state-of-the-art and future molecular gas surveys.  In $\S$~\ref{sec:appM51}, we provide an explicit example of
how we estimate GMC lifetimes in M51 by applying our method to the
PAWS data.

\subsection{Clouds in observations and in theory}
\label{sec:cloudsObs}
The application of our method to actual observational data
involves an obvious but important assumption, i.e. that a galaxy's GMC
population is accurately described by the cloud catalog constructed
from observations. In practice, this means that the cloud lifetime strictly applies only to
objects that are well-represented by the observational data, which depends on observational characteristics such as the resolution and sensitivity of the survey.  

For example, the PAWS data that we analyze in $\S$~\ref{sec:appM51} has a
spatial resolution of $\sim40$\,pc and a spectral resolution of 5\,km $s^{-1}$,
with an RMS noise of 0.4\,K per channel (\citealt{petyPAWS}; \citealt{schinner2013}). Assuming a Galactic conversion factor from CO luminosity to molecular gas mass, a 10$^5$ $M_\odot$
GMC in M51 should therefore be detected by the PAWS survey with
5$\sigma$ significance.  This then translates into a completeness limit for the cloud catalog constructed from these (or similar) observations through decomposition of the CO emission, i.e. with CPROPS (see Colombo et al. 2014).  
Our definition of `cloud' thus applies only to emission above the sensitivity limit when this emission is moreover connected in position-position-velocity ({\it ppv}) space (as assumed in the decomposition; see $\S$~\ref{sec:cloudcatalog}).    

\subsubsection{The cycling of molecular material}
Some part of the CO emission detected in galaxies may not be in cloud form at all, as highlighted by PAWS \citep{petyPAWS}.  Roughly half of all CO emission surveyed throughout the PAWS field of view is composed of  clouds (dropping to $\sim$40\% in the inter-arm; \citealt{colombo2014a}).  The remaining CO emission, which we henceforth refer to as part of the `non-cloud medium', is partly resolved, but not organized into coherent `cloud-like' entities, and partly structured on much larger scales (\citealt{colombo2014a}; \citealt{petyPAWS}). 
Regardless of the nature of this emission, when clouds evolve outside the definition of a cloud catalog we assume that they `die': the measured lifetime marks the time either when clouds are dispersed back in to the non-cloud medium or when their masses and sizes fall below our detection threshold.\footnote{We note that the total emission in the cube does not support the presence of significant numbers of clouds below 10$^5$$M_\odot$, extrapolated from the low-mass end of the mass spectra of cataloged clouds \citep{colombo2014a}.}  
Similarly, we assume that a cloud is not formed until it grows beyond the detection threshold of the dataset.  %mass $M$=10$^5$$M_\odot$ or size $a$$>$40pc.  

\subsubsection{Phase changes and continuity}
\label{sec:continuity}
In general, clouds could also `disappear' as a result of phase changes.  Whether clouds cataloged with a particular tracer represent a complete sample (and thus supply a useful tracer of cloud lifetimes in an Eulerian representation) depends on whether the gas phase being traced undergoes transitions to other phases (as considered in the case of M51 in section $\S$ \ref{sec:contM51}).  

Where the ISM is molecule-dominated, as in the central area of M51 covered by the PAWS field of view, we can use CO as our primary tracer of the gas including its cloud entities.  
In other instances (in other galaxies with different balances in their atomic and molecular phases), variation in the mass in a given phase from one side of the inter-arm to the other may imply that additional gas phases must be accounted for, or that the CO-to-H$_2$ conversion factor varies  azimuthally.   Before applying our method, we therefore recommend first assessing mass continuity across the inter-arm as considered later in $\S$ \ref{sec:contM51}.  Note, though, that continuity is not required by our method, even if it may provide a useful diagnostic; see $\S$ \ref{sec:strengths}.  %Our method is also independent of mass estimation, and so is free of the related uncertainties.  

\subsection{(Relative) Strengths and weaknesses of the method}
\label{sec:strengths}
Although the lifetime we measure is arguably sensitive to the identification and decomposition that defines clouds (as discussed in $\S$ \ref{sec:cloudsObs}; requiring, in particular, coherent structures in {\it ppv} space%\footnote{Here we associate incoherence in {\it ppv} space with the absence of a 'true' cloud, rather than a failure of the decomposition or catalog incompleteness. })
, it is independent of the actual cloud mass (and how it varies with time).  We therefore avoid several of the uncertainties associated with mass estimation, from the definition of the cloud boundary itself to the conversion of the surface brightness of a given tracer to gas column density (e.g. the CO-to-H$_2$ conversion factor).   

\subsubsection{Relation to star formation}
The method is also independent of the presence of star formation, which can often be difficult to determine (due to strong dust extinction or sensitivity limitations).  Our cloud lifetime measurement thus avoids such ambiguities and is neither limited to the onset of star formation nor sensitive to the duration of star formation.  In principle, the measured lifetime can extend beyond the star formation event, as long as the cloud remains more massive than 10$^5$M$_{\odot}$.  

\subsubsection{Population Growth}
Our method is less powerful for strictly estimating cloud lifetimes when the cloud population is also characterized by growth due to, e.g. cloud formation or transformation, which may often accompany star formation (and associated feedback). When clouds are introduced into the population between zone I and zone II, the measured $F_{lost}$ will underestimate the true fraction of `lost' clouds.  Eq. (\ref{eq:numlost}) thus in general provides only an upper limit on the cloud lifetime, unless the number of clouds added to the population is far exceeded by the number of clouds lost (i.e. the timescale $\tau_{grow}$ to gain clouds is much longer than the true cloud lifetime $\tau_{true}$).  

Under most circumstances in which the (inter-arm) population contains growth, we expect the timescales for population growth and loss to be similar, i.e. clouds emerge from the destruction or transformation of existing clouds.  When comparable numbers of clouds are gained and lost between zone I and zone II, there will be no evolution in cloud number.  As $F_{lost}$ approaches zero, $\tau$ will deviate strongly from the true cloud lifetime.\footnote{For populations predominantly undergoing growth (with comparatively few losses) $F_{lost}$$<$0, and eq. (\ref{eq:numlost}) naturally yields a lower bound on the timescale with which clouds are formed.}  In this scenario, eqs. (\ref{eq:numlost}) and (\ref{eq:timescales}) can provide an estimate for the population growth timescale, assuming a specific cloud lifetime.   

\subsubsection{Other assumptions of the method}
Our assumption that the rates of gain and loss in the population are independent of time is one of necessity, but it should not be entirely unrealistic, e.g. when describing losses due to shear. (Changes in the galaxy's mass distribution and hence orbital velocities will typically vary over much longer timescales than an inter-arm travel time.)  For other mechanisms, both the rate itself and how quickly it changes could depend on cloud properties, such as mass or surface density; if the process of star formation is limited to clouds of a certain type, then destruction via feedback will only act on this subset of the population.  With a large enough sample of clouds, it should be possible to separate clouds into subpopulations in order to identify whether such a dependence might exist. 

While these factors can be easily integrated into alternative approaches, e.g.  in which cloud lifetimes are estimated by fitting tailored models mixing cloud formation and destruction to cloud mass spectra, our simple (reductive) approach affords relative model-independence, requiring only measurements based on observables..pdf

\begin{figure*}[t]
\begin{center}
\begin{tabular}{c}
\includegraphics[width=.95\linewidth]{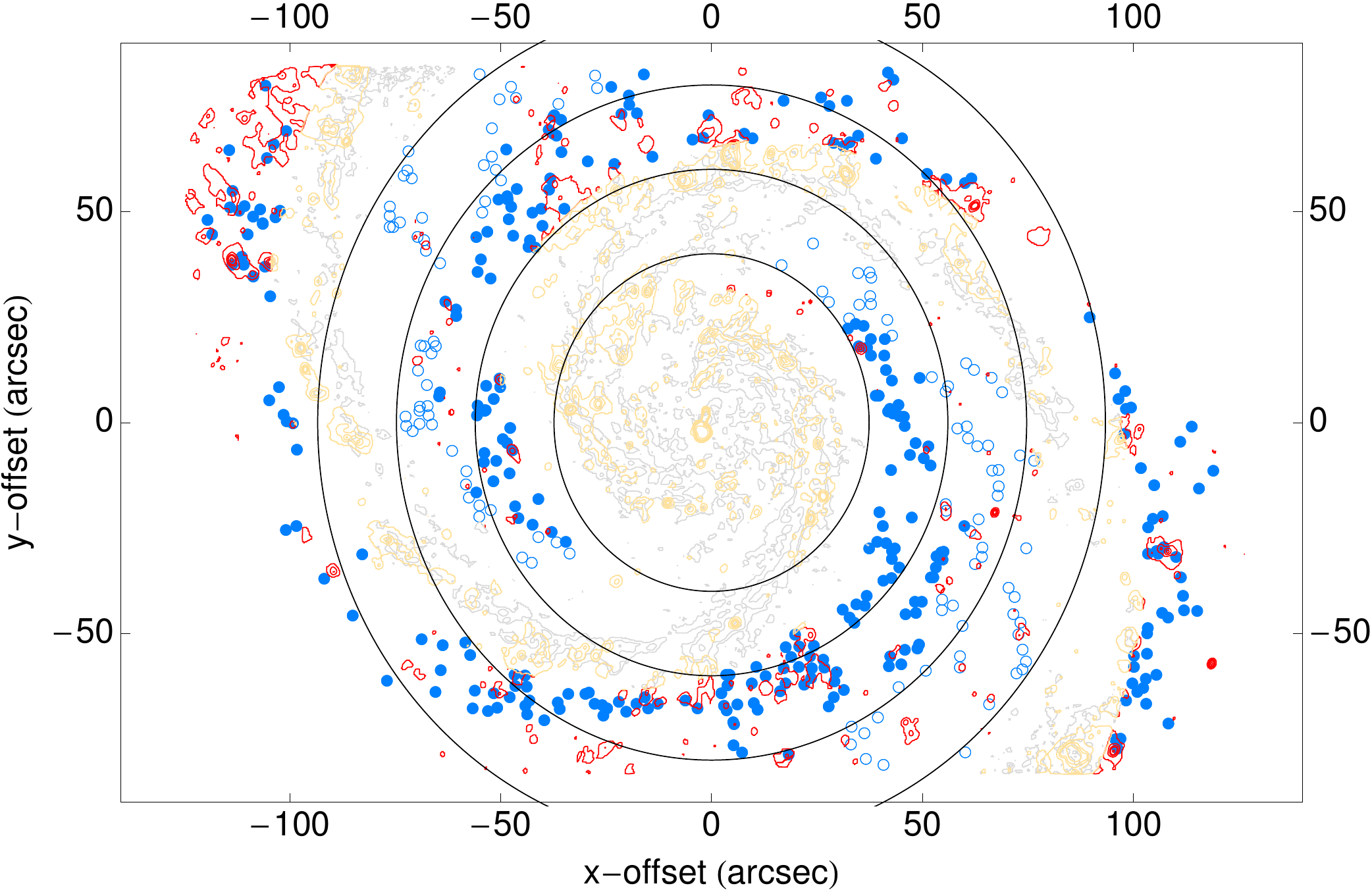}
\end{tabular}
\caption{Map of inter-arm cloud positions (blue) extracted from the PAWS survey of CO(1-0) emission in M51 by \citet{colombo2014a}. Clouds in zone I of the inter-arm are marked with filled circles, while zone II clouds are shown with open circles.  Gray contours highlight the position of the spiral arms traced in CO. Contours of H$\alpha$ emission across the PAWS FOV (from \citealt{schinner2013}; assuming the \citet{Gutierrez} stellar continuum correction of the HST ACS image) are shown color-coded by environment: the zones downstream and upstream in the inter-arm (red) and the arm and center environments (yellow).  Concentric black rings mark radii $R$=40", 60", 80" and 100".  
}
\label{fig:map}
\end{center}
\end{figure*}
\section{Application to M51}
\label{sec:appM51}
\subsection{Cloud-decomposed inter-arm emission}
\label{sec:cloudcatalog}
Our estimate for the cloud lifetime in M51's inter-arm relies
on the catalog of GMCs identified in PAWS \citep{colombo2014a}. This
catalog contains $\sim1500$ clouds, and includes measurements of the
cloud position, size, linewidth, peak brightness, integrated CO
luminosity and dynamical mass. Along with the spiral arms, molecular
ring and nuclear bar (see \citealt{colombo2014a}), the inter-arm is one
of four main dynamical environments in M51's inner disk. Just over 500 clouds populate this region in the disk, defined as the area between the environment of the two main spiral arms\footnote{The spiral arms have a finite, kinematically-determined angular width \citep{colombo2014a}.} and extending from $R$=1.3 kpc, from the `center environment', to the edge of the field of view (see Figure \ref{fig:map}).  This large number of clouds
($N_{clouds}>500$) is required to accurately monitor changes in the
cloud number density as a function of azimuth and galactocentric
radius within the inter-arm environment.

Inter-arm clouds are split into two populations, one on either half of the inter-arm as defined by Colombo et al. (2014a).  The inter-arm sits between either of the two main spiral arms, whose widths are determined via observed gas kinematics as is, therefore, the location of the inter-arm mid-point itself.   The distinction between zone I and zone II is illustrated in Figure \ref{fig:newpolar} showing the polar-coordinate projection of clouds in the inter-arm to the south relative to the two spiral arms.  Although the particular sorting of clouds into zone I and zone II can be sensitive to the exact location of the inter-arm midpoint, quantifying our uncertainty in this location serves as a way of evaluating the dominant uncertainty on cloud lifetimes measured with our technique (as defined in section 4.1).  As there are two main spiral arms in M51, there are are two inter-arm regions.  Following the sorting of clouds into zone I and zone II within each inter-arm (into four populations), their number densities are determined and then combined together (into two main populations).     

According to \cite{colombo2014a}, the 5$\sigma$ sensitivity of the PAWS dataset is $1.2\times10^{5}$\,$M_{\odot}$ throughout most of the field of view, but an increase in the noise toward the edge of the field increases the completeness limit to $3.6\times10^{5}$\,$M_{\odot}$.  We therefore choose to include all cataloged clouds, which extend down to $1.2\times10^{5}$\,$M_{\odot}$, in our analysis at all radii except at $R$$>$70", where we keep only clouds more massive than $3.6\times10^{5}$\,$M_{\odot}$ given the change in noise pattern.  We find very little change in our results with small changes in this threshold, as most clouds are well above this mass.  

\subsection{Cloud trajectories} 
\label{sec:cloudtrajectories}
\begin{figure}[t]
\begin{center}
\begin{tabular}{c}
\includegraphics[width=.95\linewidth]{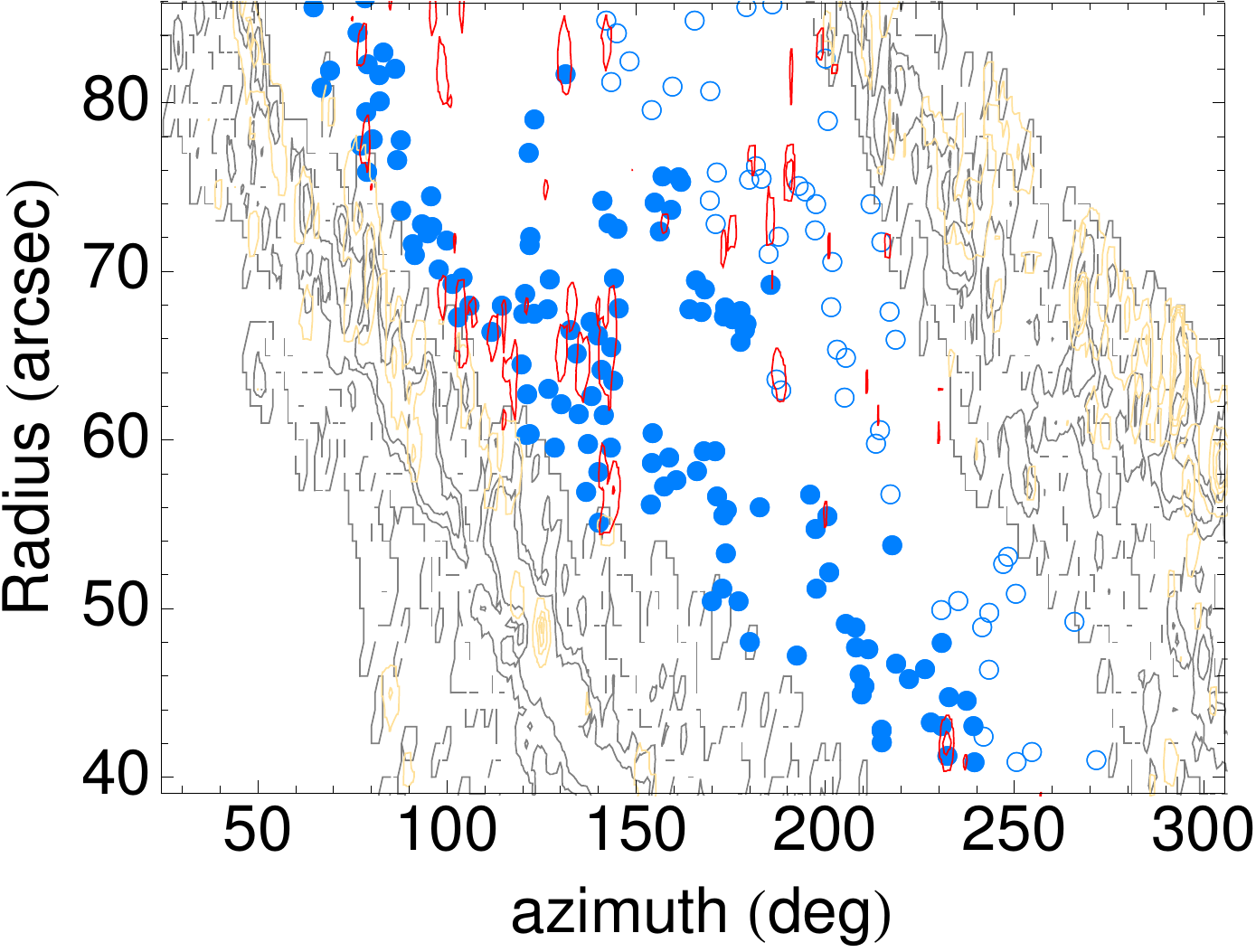}
\end{tabular}
\caption{Polar coordinate representation of Fig. \ref{fig:map}, highlighting clouds in the inter-arm to the south.  Clouds in zone I are marked with blue filled circles, while zone II clouds are shown with blue open circles.  As in Fig. \ref{fig:map}, gray contours highlight the position of the spiral arms traced in CO, while contours of H$\alpha$ emission are shown color-coded by environment: the zones downstream and upstream in the inter-arm (red) and the arm and center environments (yellow).  
}
\label{fig:newpolar}
\end{center}
\end{figure}

According to our definition of the inter-arm environments, clouds leave the spiral arm downstream and enter zone I and then proceed to zone II located upstream of the next spiral arm.  The trajectory of clouds during their inter-arm passage is expected to be roughly circular (e.g. straight left to right in Figure \ref{fig:newpolar}), but it is important to verify that clouds do not drift significantly in galactocentric radius.  
In the event of large radial excursions, changes in cloud
numbers from zone I to zone II could be explained entirely as clouds
passing from zone I in one radial bin to zone II in the neighboring
bin, or even passing directly between radial bins within zone I
itself.   

In general, clouds in spiral galaxies are subject to large non-circular motions as they orbit in the non-axisymmetric potential.  
But once a cloud exits the spiral arm, where these motions are largest, the cloud undergoes nearly circular motion in the inter-arm (i.e. \citealt{roberts69}).  
We can estimate a cloud's radial
excursion about its circular orbit as roughly the size of the epicyclic radius, $v_r/\kappa$, where $\kappa$ is the epicyclic frequency and $v_r$ is the size of radial streaming motions.  These motions are modest in the inter-arm of M51 compared to the spiral arms.  
But we can place a conservative estimate on the radial distance a cloud will
traverse during its inter-arm passage by adopting the maximum value of radial motions in the spiral arm (30 km s$^{-1}$; \citealt{meidt2013} and \citealt{colombo2014b}) 
together with the radially varying
$\kappa$ obtained by \citet{meidt2013}.  
We find that the radial excursion is everywhere only $\sim$ 4\% of the orbit circumference (even at a radius $R$=4 kpc, this is at most 300 pc).  Clouds can thus be safely assumed to follow circular paths as they cross the inter-arm environment.   

\subsection{Travel time in the inter-arm} 
\label{sec:traveltime}
Given that clouds in the inter-arm follow circular paths, the time it takes for a test cloud to travel the current distance spanned between any two of $m$ spiral arms can be estimated as $t_{travel}$=$(2\pi-m\theta_{arm}) R m^{-1}V_{iarm}^{-1}$, where $\theta_{arm}$ is the angular width of the spiral arm and $V_{iarm}$ is the azimuthal inter-arm velocity.  In M51 $m$=2, since there are two spiral arms, and we approximate $V_{iarm}$$\approx$$V_{rot}$ (i.e. little inter-arm streaming) so that $t_{travel}$=$\pi R/V_{rot}$.  This travel time is good to within $\sim$ 4 Myr, 
accounting for the assumed finite, uniform spiral arm width $w$=$R\theta_{arm}$$\sim$300 pc.  Note that although we write $t_{orb}$=$2\pi R V_{rot}$=2$t_{travel}$ throughout, this underestimates the true time to complete one (non-circular) orbit, which is $2(t_{arm}$+$t_{travel}$) with $t_{arm}$=$w V_{arm}^{-1}$, where $V_{arm}$ represents transverse motions through the spiral arm potential that reach $\sim$ 30 km s$^{-1}$ \citep{meidt2013}.  
As we are interested only in the time to travel the distance spanned by the two spiral arms (and its uncertainty) at each radius we simply adopt the circular velocity from the rotation curve model derived by \citet{meidt2013} from the baryonic mass distribution in M51.  

\begin{figure}[t]
\begin{center}
\vspace*{-.4in}\includegraphics[width=1.15\linewidth]{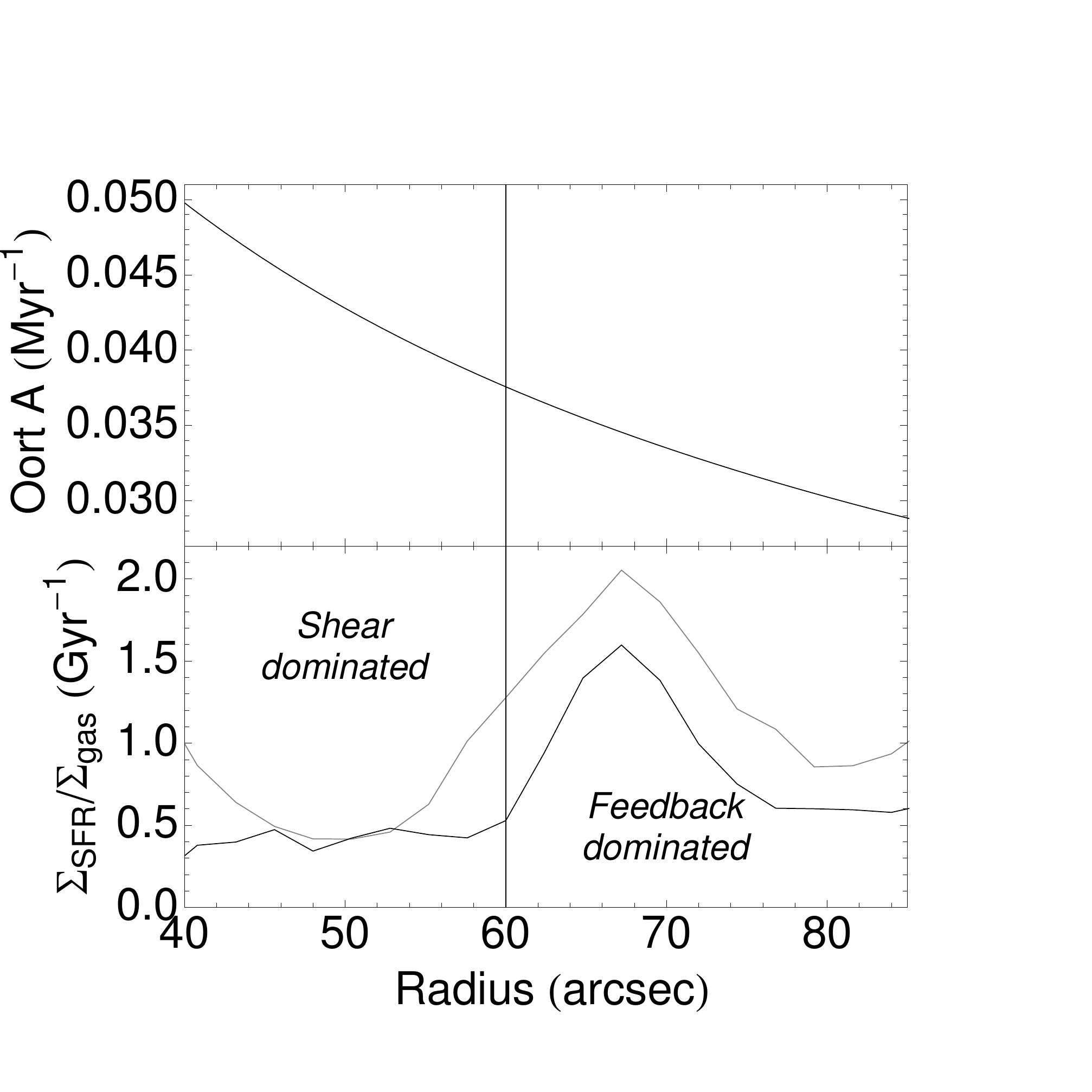}
\vspace*{-.3in}
\caption{(Top) Radial profile of inter-arm shear measured by the background Oort A estimated in M51 by \citet{meidt2013}. (Bottom)  Radial trends in $\Sigma_{SFR}$/$\Sigma_{gas}$ (inverse gas depletion time $\tau_{dep}$), as measured by \citet{meidt2013}.  Measurements are extracted in radial bins in which azimuthal averaging of $\Sigma_{gas}$ (estimated from PAWS CO and THINGS HI) and $\Sigma_{SFR}$ (measured from H$\alpha$ and 24 $\mu m$) runs from 0 to 2$\pi$ (throughout the PAWS FOV; gray) or only across the inter-arm (black).  Uncertainties are on the order of 30\% \citep{meidt2013}.  We interpret regions at $R$$>$60" where the star formation rate per unit gas mass is high as locations where feedback from massive star formation dominates cloud destruction.  The zone in which the star formation is reduced but shear is relatively high marks the region where we expect cloud destruction primarly through shear. }
\label{fig:timescales}
\end{center}
\end{figure}
\subsection{Radial variation in cloud formation/destruction}
M51 presents a unique opportunity to distinguish between two main cloud dispersal processes, i.e. shear and feedback from massive star formation.  
Over the radial range spanned by the PAWS field-of-view,
our knowledge of the galaxy's mass distribution, gas kinematics and
global pattern of star formation indicates that there are two distinct
radial zones where we can expect the influence of each of these
processes to dominate. 
As shown in Figure \ref{fig:map}, most of the massive star formation in M51 occurs along the spiral arms, limited to radii $R$$>$60''\footnote{The evident suppression of star formation along the inner spiral segment, between 40''$<$$R$$<$60'', is discussed by \citet{meidt2013}.}, and appears offset just downstream of the spiral arms traced at high resolution in the PAWS CO(1-0) map.  The majority of this star formation falls within our definition of the arm environment, but it persists into zone I of the inter-arm.  Clouds at these galactocentric radii thus appear 
susceptible to feedback, either from star formation
that is internal to the clouds themselves or from star formation
activity in the nearby spiral arm.

At smaller radii ($R$$<$60'') there is relatively less massive star formation, and here the impact of shear and Coriolis forces due to disk differential rotation are better highlighted.  
These factors can lead to cloud dispersal and destruction at all radii in the inter-arm, but shear measured by the background Oort $A$ is notably larger here than beyond $R$$\sim$70'', as shown in the top panel of Figure \ref{fig:timescales}.  This measure of shear should be appropriate in the inter-arm, where we expect non-circular streaming motions to be negligible.\footnote{Note that in the spiral arms where streaming motions are larger, shear described by Oort A accounting for these motions (and including the background) can behave very differently, suggesting that the impact of shear on cloud stability may differ between the inter-arm and the arm. }

In the bottom panel of Figure \ref{fig:timescales} we also plot the radial profile of the ratio of the star formation rate surface density to the molecular gas surface density in the inter-arm for reference.  Beyond $R$=60", the high star formation per unit gas mass suggests that feedback may be more disruptive here than at smaller radii where, in contrast, shear is high.  In what follows, we therefore
distinguish between the ``shear-dominated'' and ``feedback-dominated''
regions of M51's inner disk, which we define as being galactocentric
radii 41''$<$$R$$<$60'' and 60''$<$$R$$<$91'' respectively.

\subsection{Phase changes and continuity in M51}
\label{sec:contM51}
\begin{figure}[t]
\begin{center}
\vspace*{-.2in}\includegraphics[width=1.\linewidth]{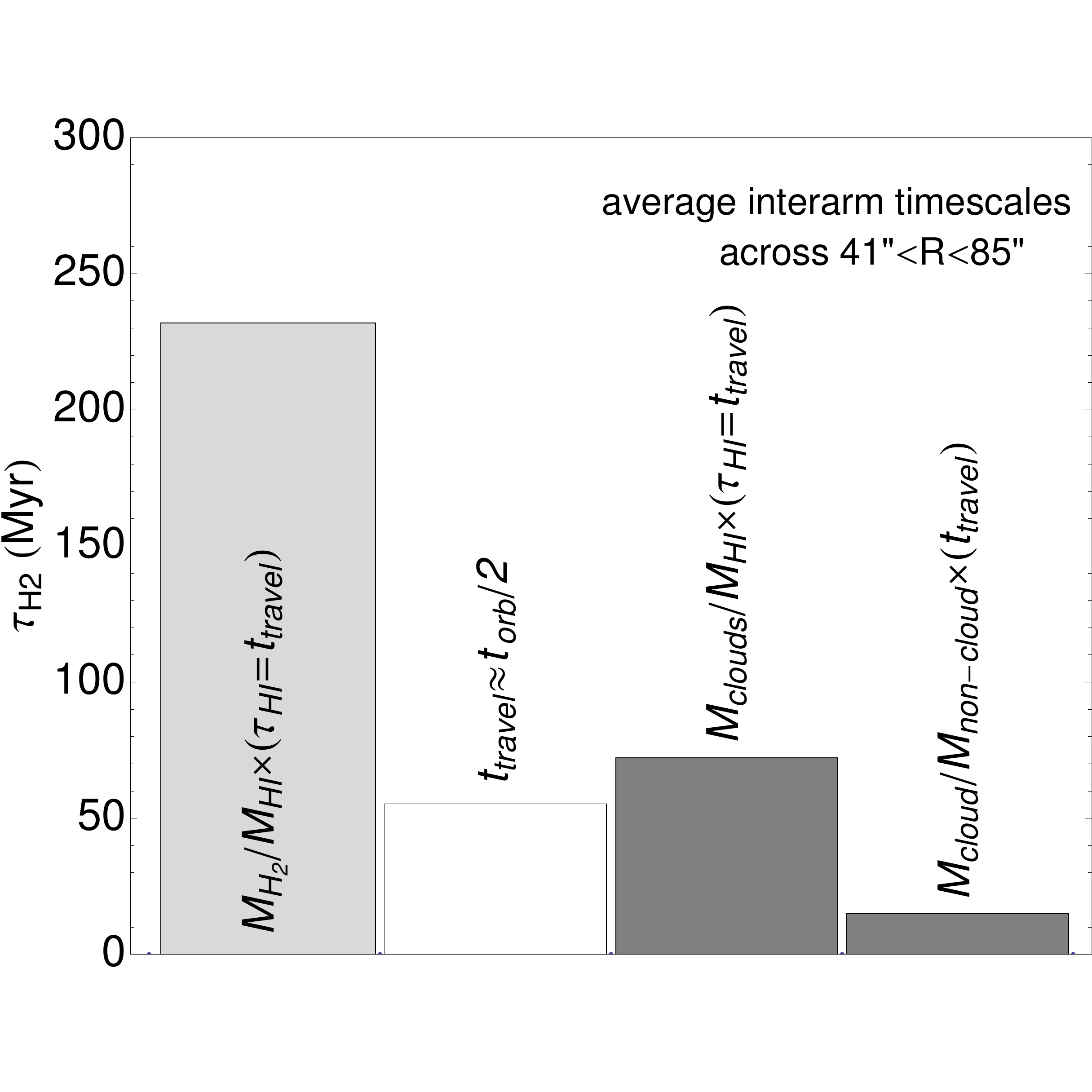}
\caption{Histogram of the average characteristic timescales for molecular hydrogen in the inter-arm of M51 from throughout the zone 41''$<$$R$$<$91''.  The light gray bar shows the time for the ISM to remain molecular given conversion to atomic hydrogen, which is assumed to have characteristic lifetime $\tau_{HI}$ set by half the dynamical timescale ($t_{travel}$, the travel time from one of two spiral arms to the next; shown here as the white bar).  The set of dark gray bars show the characteristic timescale for the part of the molecular hydrogen in cloud form, as cataloged by PAWS, assuming conversion either directly to atomic hydrogen or back to the non-cloud medium, including molecular hydrogen.  Radial profiles of the latter two timescales are shown in Figure \ref{fig:both}.}
\label{fig:contTimescales}
\end{center}
\end{figure}

\begin{figure}[t]
\begin{center}
\vspace*{-.3in}\includegraphics[width=1.15\linewidth]{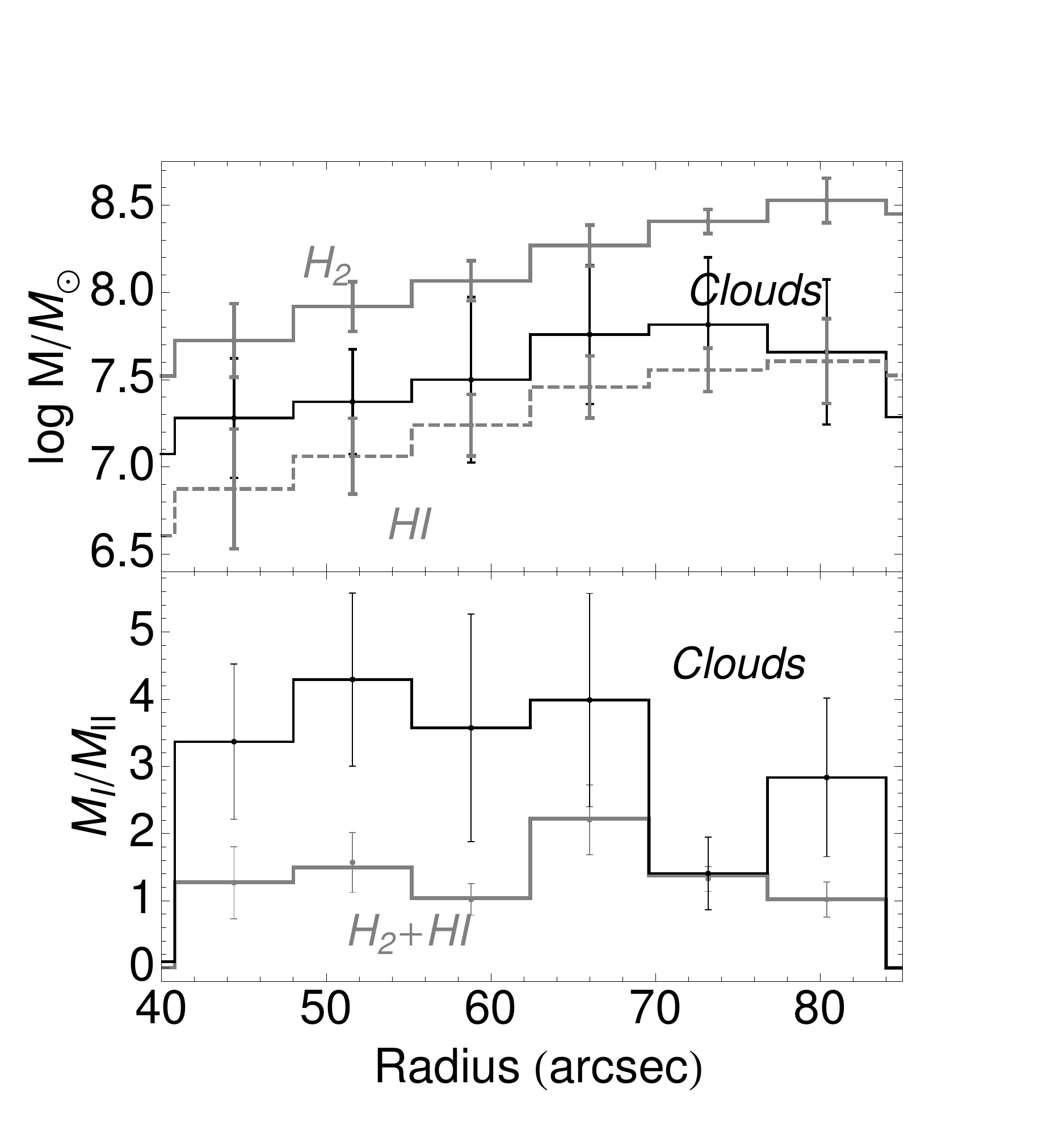}
\caption{(Top) Radial profiles of the mass in molecular gas (gray), atomic gas (gray dashed) and molecular clouds (black) in the inter-arm of M51.  Measurements and error bars adopt the scheme used later in $\S$\ref{sec:measUnc}, i.e. radially binned averages for which the uncertainties arise with modifications to the bin definition.  Errors on the mass in clouds additionally include (and are dominated by) propagated measurement uncertainties on individual cloud masses.  (Bottom) Ratio of the total (molecular plus atomic) gas mass in zone I compared to zone II (gray) and the ratio of the total mass in clouds in zone I compared to zone II (black).}
\label{fig:contprofiles}
\end{center}
\end{figure}

%\ref{sec:continuity}
In Figures \ref{fig:contTimescales} and \ref{fig:contprofiles} we confirm that CO emission supplies a complete picture of the evolution of M51's inter-arm molecular cloud population, i.e. that phase changes do not cause incompleteness in the PAWS catalog and the transition from molecular gas to atomic gas occurs over a much longer timescale than the orbital period (and the expected cloud lifetime). The left-most timescale in Fig. \ref{fig:contTimescales} in particular represents the length of time over which the ISM is expected to remain molecular, calculated based on the requirement of continuity between the atomic and molecular phases (i.e. the two dominant ISM
components in M51 by mass).  

Following \citet{scoville} (and \citealt{koda09}), $\tau_{H_2}=\tau_{HI}M_{H_2}/M_{HI}$ where $M_{H_2}$ and $M_{HI}$ are the masses of molecular and atomic gas in the inter-arm (see Figure \ref{fig:contprofiles}), ignoring the negligible mass in ionized gas as well as conversion into stars (considering the long $>$ 1 Gyr depletion timescale implied by the current rate of star formation, $\sim$2 $M_\odot$ yr$^{-1}$; \citealt{schuster}).  
The mass of molecular hydrogen in the inter-arm, shown in the top panel of Figure \ref{fig:contprofiles}, is calculated from the PAWS CO emission assuming a CO-to-H$_2$ conversion factor of $X$=2$\times$10$^{20}$ cm$^{-2}$ (K km s$^{-1}$)$^{-1}$, which has been found to apply across the PAWS field-of-view (\citealt{colombo2014a} and references therein).  The HI mass (also shown in Figure \ref{fig:contprofiles}) is estimated using the THINGS survey data \citep{walter} assuming that the HI emission is optically thin.    

According to \citet{scoville}, the HI lifetime $\tau_{HI}$ can be approximated by the shorter of either the dynamical time ($\sim$$2t_{travel}$ here; see $\S$ \ref{sec:appM51}) or the spiral arm passage time $t_{sp}=2\pi/m(\Omega-\Omega_p)$.  We adopt the former as it is shorter than $t_{sp}$ between the $m$=2 spiral arms in M51, assuming the spiral pattern speed $\Omega_p$ estimated by Querejeta et al. 2015 (in prep; see also Meidt et al. 2013).\footnote{The spiral arm passage time would apply in the case that spiral arm passage prompts phase changes from atomic to molecular gas (i.e. \citealt{dobbs08}; \citealt{VS07}) and this happens faster than a dynamical time (i.e. \citealt{scovilleWilson})}. We obtain a more conservative estimate for $\tau_{H2}$ by letting $\tau_{HI}$=$t_{travel}$ as it everywhere underestimates the cycling timescale in the case where $\tau_{HI}$=$t_{dyn}\sim2t_{travel}$.  This shorter timescale also allows for direct comparison between the estimated $\tau_{H2}$ and the time window $t_{travel}$ probed by the azimuthal span of our analysis region in the present inter-arm.  

The fact that the left bar depicted in Figure \ref{fig:contTimescales} greatly exceeds $t_{travel}$ therefore immediately suggests that the disappearance of clouds from the PAWS catalog in the inter-arm is not the result of phase changes from molecular to atomic, and can instead be associated with a genuine finite cloud lifetime.  
As noted in $\S$ \ref{sec:continuity}, other galaxies may have lower molecular gas fractions, in which case the implied length of time that the gas stays in molecular form is reduced.   Thus, we recommend first estimating the timescales described above before applying our method in order to properly assemble a multi-phase census of clouds, where necessary.  As suggested in $\S$ \ref{sec:continuity}, we also recommend assessing mass continuity across the inter-arm as demonstrated in the bottom panel of Figure \ref{fig:contprofiles} in order to insure that these timescales are meaningful.  

In M51, H$_2$ traced by CO dominates the total gas mass, which is almost equally distributed between zone I and zone II at all radii in the inter-arm.  This suggests that the molecular phase largely captures mass continuity.  The large variations in the mass in cloud form from one side of the inter-arm to the other shown in the bottom panel of Figure \ref{fig:contprofiles} therefore imply that clouds leave the population as a result of destruction rather than phase changes.  

Note that the above continuity argument also suggests that inter-arm clouds are converted directly back to the atomic phase only very slowly, with a timescale which is again long compared to $t_{travel}$ (third bar from the left), assuming once again $\tau_{HI}$=$t_{travel}$.   This long timescale for conversion to atomic hydrogen implies that molecular clouds in M51 would primarily evolve quickly back into their molecular hydrogen surroundings.  

The same continuity argument indeed implies a very fast conversion between clouds and the surrounding molecular gas (right bar).  As noted earlier in $\S$ \ref{sec:cloudsObs}, the cloud component of the inter-arm molecular gas in M51 (comprised of objects more massive than 10$^5$ M$_\odot$) represents $\sim$ 40\% of the total inter-arm CO flux mapped by PAWS (see also Figure \ref{fig:contprofiles}).  This once again assumes that $t_{travel}$ is the characteristic time for the molecular (and atomic) gas to remain `outside' clouds (i.e. assuming the medium is converted into clouds during passage through the spiral arm), and that the same CO-to-H$_2$ conversion factor applies to cloud and non-cloud CO emission traced by PAWS \citep{liszt}.  
Such a short cloud lifetime by this measure is consistent with the lifetimes found here (presented in $\S$ \ref{sec:measurements}), which also captures cloud conversion back to the (molecular) non-cloud medium.  But our estimate requires no prior knowledge of the timescale characteristic of the non-cloud medium, which may deviate from the assumed $t_{travel}$ depending on the true cloud formation timescale. Our method is also independent of mass estimation, and so is free of the related uncertainties, as discussed in $\S$ \ref{sec:strengths}.   


\section{Results}
\label{sec:measurements}
\begin{figure*}[t]
\begin{center}
\begin{tabular}{cc}
\includegraphics[width=.5\linewidth]{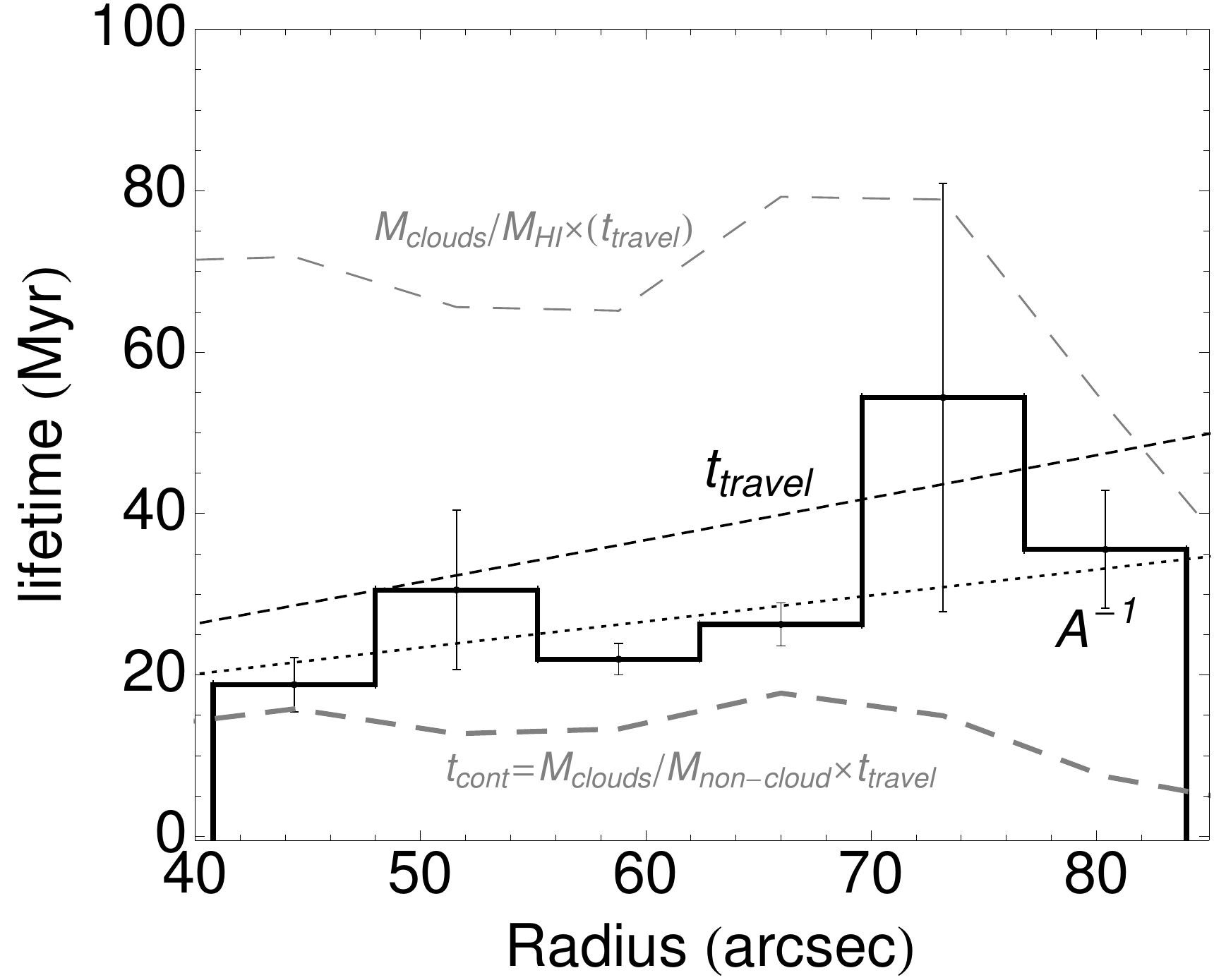}&\includegraphics[width=.5\linewidth]{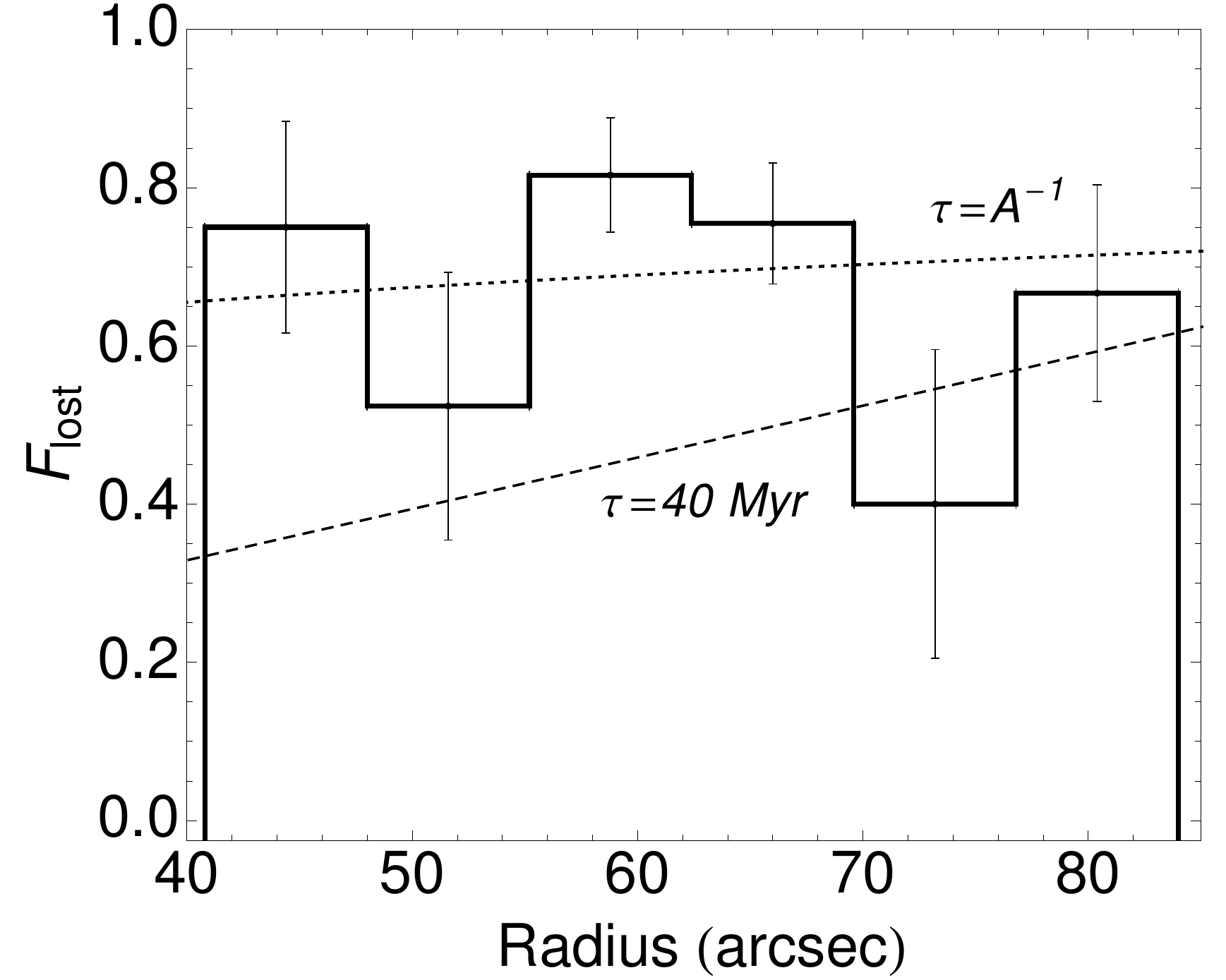}
\end{tabular}
\caption{(Left) Radial profiles of cloud lifetimes measured for the inter-arm cloud population in M51.  The solid black line shows our estimate using eq. \ref{eq:numlost} and the measurements of $F_{lost}$ shown on the right.  Error bars are propagated from the uncertainties on the measured $F_{lost}$.  The black dotted (dashed) line shows the shear timescale $A^{-1}$ (the inter-arm travel time $t_{travel}$).  The set of thick dashed gray lines show two independent cloud lifetime estimates based on continuity (see $\S$ \ref{sec:strengths}), with average values indicated by the two rightmost bars in Figure \ref{fig:contTimescales}).  (Right) Radial profile of the fraction of `lost' clouds between zone I and zone II $F_{lost}$.  Error bars represent the change in lost fraction at each radius due to our standard uncertainty (see text for details). The dotted line represents the radial behavior in $F_{lost}$ expected if the cloud lifetime is set by the shear timescale $A^{-1}$, while the dashed line assumes a fixed lifetime of 40 Myr.
Black arrows indicate the radial bin in which the number of clouds increase from downstream to upstream (see $\S$ \ref{sec:feedback}).
}
\label{fig:measurements}
\end{center}
\end{figure*}
\subsection{Measurements and uncertainties}
\label{sec:measUnc}
To best reveal the radial dependence of cloud dispersal and destruction processes expected in M51, we divide the inter-arm environment into a series of radial bins with uniform width.  Each radial bin is further split at the midpoint of its azimuthal extent into two areas, zone I and zone II, in which we count the number of clouds $N_I$ and $N_{II}$.  Radial bins are discontinued at $R$=90"=3.3 kpc, the last radius in the field-of-view where it is possible to assign equal areas to zones I and II and cloud identification is reliable.  The bin width was chosen to ensure sufficient numbers of clouds in each zone of each bin (a minimum of 4 and as many as 23) while at the same time matching our conservative estimate for the size of the maximum radial excursion expected for clouds.

The uncertainties on our measurements reflect the changes to $N_I$ and $N_{II}$ when the boundaries of the inter-arm zones are modified.  Part of the uncertainty is estimated by changing the location of radial bins by $\pm$1" (the PAWS resolution; 15\% of the bin width).  This uncertainty is added in quadrature with the error arising from displacement of the inter-arm midpoint by 8$^\circ$, accounting for the $\sim$ 4 Myr uncertainty in our estimate of the inter-arm travel time.  

For reference, with this standard error budget the uncertainty on the total gas mass measured in either zone I or zone II is 30\% \citep{colombo2014a}, comparable to the fraction of mass in cloud form.  The significance of the mass-based diagnostics that include this error, which we use later in $\S$ \ref{sec:fate} to interpret our measurements of $F_{lost}$ and $\tau$, is therefore limited.  However, the significance of our measurements of $F_{lost}$ and $\tau$ are not compromised.   

\subsection{Evolution in M51s inter-arm cloud population}
In Figure \ref{fig:measurements} we present our measurements of
$F_{lost}$ (right) and the cloud lifetime implied by this evolution in
cloud numbers (left) for the inter-arm region of M51 probed by PAWS.  Already by mid-way through the inter-arm almost 80\% of the population has been destroyed, implying a very short lifetime.  Inside $R\sim70$'',
our method yields a characteristic cloud lifetime of only 20 to
30\,Myr. 
Furthermore, we find only modest variation in
the evolution of clouds numbers from zone I to zone II with galactocentric
radius. 
This is in contrast to the increasing losses expected in a population with a uniform cloud lifetime as the travel time lengthens with radius.  Instead, the trend on the right is qualitatively consistent with cloud dispersal primarily through shear, which weakens with galactocentric radius and thus leads to longer lifetimes at larger radii (in the absence of other cloud destruction mechanisms).  

Quantitatively, moreover, we find that our lifetime
estimate agrees very well with the shear timescale (especially at
$R<$70''; left), strongly suggesting that shear is the primary
mechanism responsible for the finite lifetimes of clouds in M51's
inner disk.  While the data are consistent with the shear trend over the range 40"$<$$R$$<$85" within 1$\sigma$, we can reject the constant average value of 31 Myr with 3$\sigma$ confidence.  
Note that even at the smallest radii where the inter-arm travel time is shortest, few clouds appear to survive in the presence of such strong shear.  

At larger radii ($R>60$''), though, Figure~\ref{fig:measurements} shows that the good agreement between
the cloud lifetime and the shear timescale breaks down.  At these larger radii, shear is slightly weaker and maps of star formation rate tracers (e.g. H$\alpha$, infrared, UV) suggest that
star formation (and feedback) may have a greater influence on cloud
evolution.  
Between 70"$<$$R$$<$85", the cloud number density once again decreases
from zone I to zone II, but not as many clouds are lost as would be expected under the influence of strong shear (i.e. $F_{lost}$ falls below the shear prediction; right panel Figure \ref{fig:both}).  The cloud lifetime
estimated by our method thus appears lengthened in comparison with the shear timescale (if only marginally).  
These trends are consistent with growth in the population between zone I and zone II at the same time as clouds are destroyed.  Our measurements may 
therefore suggest that cloud formation and/or transformation become
increasingly important beyond $R>60$''.  
As discussed in $\S$\ref{sec:framework}, our
method provides only an upper limit on the cloud lifetime unless the
number of clouds added to the population is negligible compared to the
number of clouds that are destroyed.  Hence, our conversion
from $F_{lost}$ in the right panel of Figure~\ref{fig:measurements} to the
cloud lifetime shown in the left panel 
becomes highly uncertain at larger radii.  In the next section, we
examine the fate of clouds within M51's inner disk in more detail,
examining how their mode of destruction (shear or feedback) affects
their fate (i.e. fragmentation/dispersal versus transformation) and
the lifetime that we estimate using our method.

\section{Interpretation}
\label{sec:interpretation}
\subsection{The fate of clouds}
\label{sec:fate}
\begin{figure}[t]
\begin{center}
\includegraphics[width=1.15\linewidth]{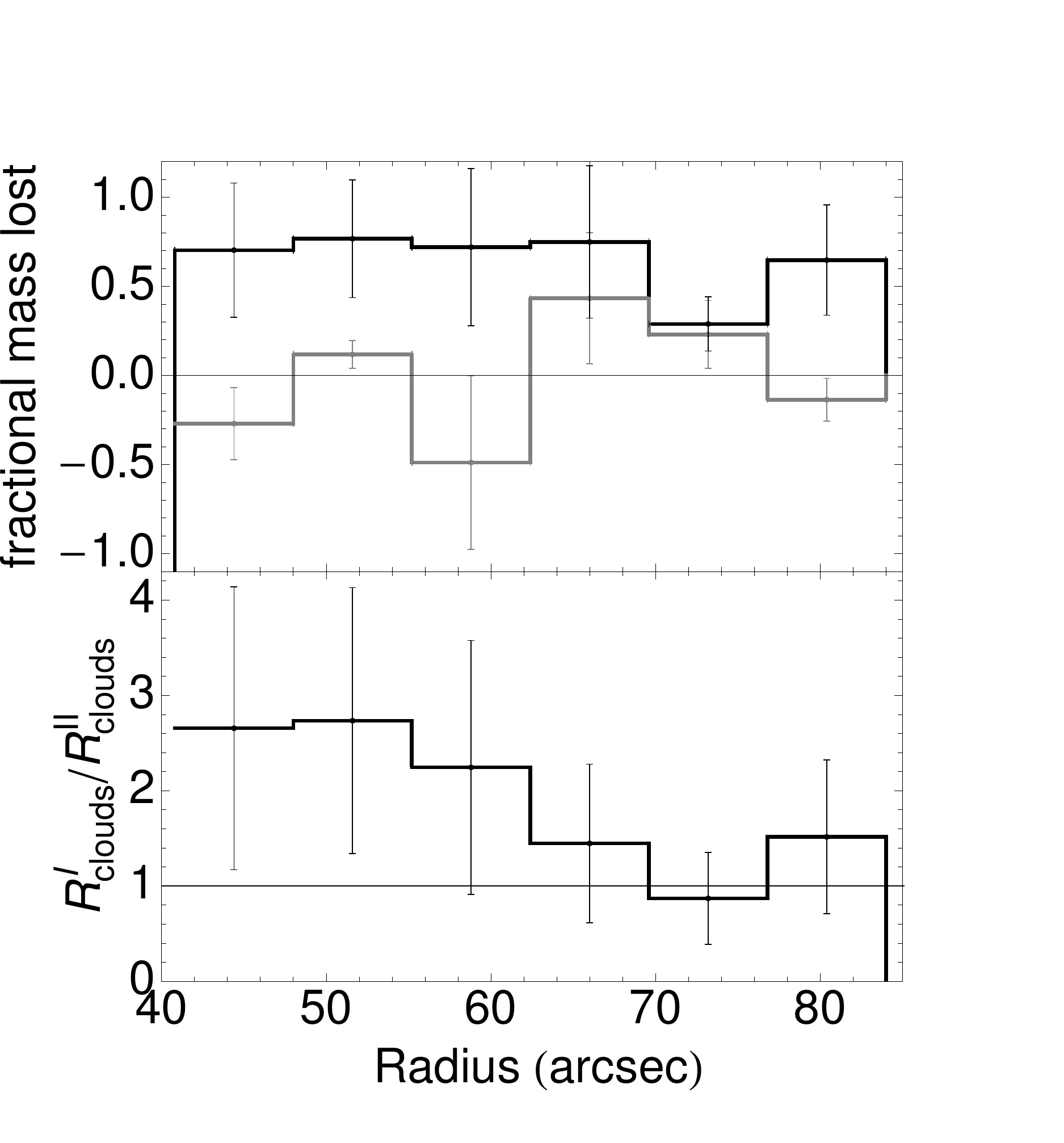}
\caption{(Top) Radial profiles of the fractional mass in clouds lost from zone I to zone II in the inter-arm of M51 (black) and the fractional mass in the non-cloud medium lost from zone I to zone II (gray).  
(Bottom) Radial profile of the ratio of the cloud-to-total mass in zone I $R_{clouds}^I$ compared to the cloud-to-total mass in zone II $R_{clouds}^{II}$ (see text for details).  }
\label{fig:both}
\end{center}
\end{figure}

Figure \ref{fig:both} shows two additional diagnostic measures of cloud evolution.   In the top panel we show variation in the mass in clouds (black) and the mass outside clouds (gray) from zone I to zone II, calculated as $(M_{I}-M_{II})/M_{I}$.  In the former case, the cloud mass in either zone $i$ is measured as the sum of the masses of $N_i$ individual clouds.  In the latter case, the non-cloud mass in either zone is measured by subtracting the cloud mass from the total mass in gas in that zone, i.e.
\begin{equation}
M_{non-cloud}^i=\left(\int\Sigma_{gas}dA_i\right)-M_i 
\end{equation}
where $A_i$ is the area covered by zone $i$ and $\Sigma_{gas}$ includes both atomic and molecular hydrogen traced by THINGS and PAWS.  Note that cloud material returned to the non-cloud medium between zone I and zone II is expected to rejoin the molecular phase, as the timescale for conversion directly back to atomic hydrogen is much longer (see $\S$ \ref{sec:continuity}).  

In the bottom panel of Figure \ref{fig:both} we show the change in the cloud-to-total mass ratio $R_{clouds}$ between zone I and zone II.  Here, $R_{clouds}$ in a given zone is the total mass in clouds divided by the total neutral (molecular plus atomic) mass in gas within that area.  

Measured quantities in each radial bin are shown with error bars that include the propagated uncertainties on cloud masses tabulated in the PAWS catalog \citep{colombo2014a} and uncertainties arising from our standard error budget described in $\S$ \ref{sec:measUnc}.  While these errors can be quite large, Figure \ref{fig:both} still provides a useful basis for interpreting the trends in Figure \ref{fig:measurements}.  The two panels together suggest that clouds can undergo very different evolution depending on their radial location in the disk. 
\subsection{Shear-dominated evolution} 
\label{sec:shear}
To start, at small radii where shear is strong, there is significantly less mass in clouds in zone II compared to zone I suggesting that shear acts to disperse clouds (or cause a reduction in mass below our sensitivity limit).  The bottom panel of Figure \ref{fig:both} showing the evolution in the cloud-to-total mass ratio $R_{clouds}$ from zone I to zone II tends to confirm this scenario.  
At $R$$<$60", the mean $R_{clouds}^I$/$R_{clouds}^{II}$ is consistently (if only marginally) larger than one, suggesting that more of the total gas mass is in cloud form in zone I than in zone II.  The inverse -- more mass exterior to clouds (10$^5 M$$_\odot$ and higher) in zone II than zone I -- is consistent with the idea that shear promotes mass loss and that this mass is returned to clouds' surroundings.  

We find a hint of such an exchange between clouds and the non-cloud medium in the evolution in non-cloud mass from zone I to zone II, which tends to increase across the inter-arm at small radii (leading the gray line to drop below zero; top panel).  We caution that since the cloud contribution to the total gas mass in each zone is on the order of its uncertainty, this is only weakly revealed by our measurements.  
\subsection{Feedback-dominated evolution}
\label{sec:feedback}
The fate of clouds appears to be different at larger radii where shear is smaller and feedback has a potentially greater impact on clouds.  Between 60"$<$$R$$<$85", the loss of mass in cloud form between zone I and zone II is less than for regions at smaller galactocentric radii (Figure \ref{fig:both}, top panel).  Here as well the fraction of mass in cloud form (bottom panel) remains roughly fixed from zone I to zone II, as does the number of clouds.  This suggests a scenario in which cloud destruction is accompanied by the creation of new clouds from within the existing population, keeping the number of `lost' clouds low.  

While a small $F_{lost}$ might indicate little evolution and a genuinely long cloud lifetime, this seems less likely given that the outer zone is very clearly impacted by star formation feedback (see Figure \ref{fig:map}), which might be expected to limit cloud lifetimes below the inter-arm travel time.  Instead, if the population contains growth at a rate similar to the rate of destruction, the number of clouds would remain fixed and eq. \ref{eq:numlost} would overestimate the cloud lifetime.  Below we consider several scenarios in which cloud destruction might be balanced by cloud formation.

\subsubsection{Transformation via feedback} 
In the simplest case, 
feedback might act to split or merge clouds, transforming the cloud population, rather than completely destroying or dispersing clouds.  New clouds would thus emerge from within the existing population, as descendants of transformed clouds.  
Note that if the cloud population is undergoing transformation, Figure \ref{fig:both} suggest that clouds must be merging as well as splitting, as splitting alone would be expected to increase the number of clouds by zone II.  However, we cannot rule out that some clouds may also be completely destroyed, e.g. in the manner described in the previous section; mass loss might follow from shear, which is non-zero at these radii, or through the process of star formation itself, which might consume a large fraction of the cloud mass.  In some models, the star formation efficiency per free-fall time can be as high as 0.2-0.3 (e.g. \citealt{klessen}; \citealt{bonnell}; \citealt{VS03}; \citealt{clark05}). 

\subsubsection{Population growth}
\label{sec:growth}
It is also possible (if perhaps less likely) that new clouds at these radii are formed 
completely independently of the destruction process.  (In fact, we allow for this possibility at all radii, with the caveat that our measure of $\tau$ is an upper limit on the true lifetime.)  
To account for the trends between 60"$<$$R$$<$85" in Figure \ref{fig:measurements}, cloud creation would have to roughly balance destruction and the genuine cloud formation timescale $\tau_{grow}$ in the population would need to be comparable to the cloud lifetime.  

Although genuine population growth in the inter-arm may be surprising, cloud formation may not always be limited to the spiral arms.  In the simulations of \citealt{dobbsP13}, spirals that radially decrease in strength become less important as a site of cloud formation.  In M51, the strong two-armed spiral transitions to a weaker, material-like pattern at $R$=85" (\citealt{meidt2013}; \citealt{colombo2014b}.)  

The importance of inter-arm cloud formation might moreover be expected to increase with radius, since the time between arms is longer at larger radii.  This could provide sufficient time to form new clouds from the material that was returned to the ISM by previously destroyed inter-arm clouds.  Clouds could also form directly from the non-cloud medium, presumably at a fixed rate, thus leading our measurements to overestimate the true cloud lifetime progressively more with radius.  
To confirm whether clouds can form in the inter-arm, larger, more spatially extended molecular cloud surveys are necessary.  
At present, we take our measurements in the zone 60"$<$$R$$<$83" as most likely representing transformation within the population stimulated by feedback, as discussed in the previous section, rather than 
cloud destruction accompanied by independent population growth.  

\subsection{(In)sensitivity to level of virialization, mass and surface density}
\label{sec:sensitivity}
The strength (or weakness) of variations in the cloud population from zone I to zone II at a given radius could depend on whether all clouds experience the same limit to their lifetimes, or are 
equally susceptible to destruction.  
The stability of clouds against dispersal or destruction could depend on, e.g., cloud mass or the balance between internal kinetic energy and gravitational potential energy (level of virialization).  We might therefore expect that some clouds are never dispersed while others may evolve more rapidly than the rest of a given population.  

In M51, however, we find no significant change in average cloud properties, including the virial parameter $\alpha$ \citep{bertoldiMcKee}, cloud mass or surface density, from zone I to zone II and the two populations as a whole appear very similar (see also \citealt{colombo2014a}).  Additional tests moreover show no strong link between cloud lifetime and any given property.  Specifically, we separated clouds into subpopulations and then compared the number of clouds in zone I with a given property to the number of clouds in zone II with that same property at each radius.  The measured $F_{lost}$ and lifetime $\tau$ in each subpopulation show little significant difference, e.g. the evolution in the number of clouds with $\alpha$$<$2 is nearly indistinguishable from that of the subpopulation with $\alpha$$>$2.   
While it thus appears that no subset of clouds is more susceptible to a particular destruction mechanism than any other, we emphasize that additional splitting of the inter-arm populations in zone I and zone II likely leaves insufficient cloud statistics to reliably apply eq. (\ref{eq:numlost}).  
\section{Discussion}
\label{sec:discussion}
\subsection{Cloud destruction processes}
In the previous section we used the combination of relatively little massive star formation and strong shear at $R$$<$60", and the presence of more massive star formation at radii $R$$>$60" where shear is weaker, 
to isolate the influences of shear and star formation feedback on clouds.  We emphasize, however, that we do not directly observe cloud destruction, and thus cannot absolutely conclude which process or processes limit cloud lifetimes.  So although we do find radial variation in the way the cloud population evolves across the inter-arm -- which we interpret as the signature of two main modes of cloud destruction -- we can attribute this only generally to shear and feedback.  

Indeed, in the present study we have very little leverage on the way the two processes may work together.  The susceptibility of clouds to shear could still depend on the impact of star formation, which we observe to be non-zero even at $R$$<$60" in the inter-arm (see Figure \ref{fig:map}).  Star formation feedback could be necessary to first destabilize clouds, or it could serve to enhance the dispersal process.  Likewise, even when shear is relatively weak, it may still assist in cloud destruction through other processes, including feedback.  (Note, though, that the fate of clouds at large radii seems inconsistent with evolution driven entirely by shear, as the overall mass in clouds does not decrease from zone I to zone II; see Figure \ref{fig:both}).  

\subsection{Cloud lifetime measurements}
\label{sec:lifetimes}
Figure \ref{fig:measurements} reveals a close relation between the shear timescale measured by $A^{-1}$ and the cloud lifetime at $R$$<$60".  
This suggests that, even if star formation (and its associated feedback) is required as a prior source of cloud destabilization, when shear is strong enough it takes over and sets a more important limit on the cloud lifetime $\tau$, independent of the strength or pattern of feedback.  
The lifetimes of all clouds (at least those more massive than 10$^5$M$_\odot$ as considered here) seem equally limited by the shear timescale when shear is strong.
\footnote{In the less likely alternative scenario (introduced in $\S$\ref{sec:feedback}), in which cloud destruction is accompanied by independently forming new clouds at all radii, our measurements could imply even shorter cloud lifetimes than shown in Figure \ref{fig:measurements}. }

But as shear weakens and the shear timescale increases with galactocentric radius, at some location in the disk we expect destruction via feedback to take over and set the cloud lifetime.  According to Figure \ref{fig:both}, we would argue that feedback is accompanied by transformation (creation and destruction).  Our measurements of $\tau$ in Figure \ref{fig:measurements} thus provide only an upper bound on the cloud lifetime due to feedback and not a direct estimate.  Yet we can extract an estimate for the cloud lifetime due to feedback using the fact that the associated transformation keeps $F_{lost}$ in the population low.  
The location where this signature emerges marks the location where the lifetime due to feedback is shorter than the shear timescale.  We can thus estimate the feedback timescale from the shear timescale at this radius.  

In M51, a sustained drop in $F_{lost}$ occurs at $R$$\sim$70" where A$^{-1}$$\approx$ 30Myr, suggesting that the lifetime due to feedback is around $\tau$$\approx$30Myr.   
(This location is not surprisingly very close to the radius $R$=60" that, by construction, distinguishes between zones dominated by either shear or feedback.)
Eqs. (\ref{eq:timescales}) and (\ref{eq:numlost}) can provide an alternative estimate, assuming a particular model for how the transformation via feedback occurs (e.g. equal-mass splitting) to set the relation between the rates of cloud creation and destruction.  

\subsection{Implications of short GMC lifetimes}
\subsubsection{Cloud evolution and star formation}
The cloud lifetimes measured with our technique agree very well with the few existing observation-based estimates (typically $\tau$$\approx$20-30 Myr) made with a completely independent method (i.e. linking clouds with stellar clusters at various young ages; \citealt{bash}, \citealt{Leisawitz}; \citealt{miura}; \citealt{kawamura}).  Such short lifetimes agree with the picture of rapid cloud evolution that emerges from numerical simulations, where clouds typically have short (10-20 Myr) lifetimes (e.g. \citealt{dbp11}; \citealt{dpb12}; \citealt{dobbsP13}).  We can thus confirm that, whether in star-forming disk galaxies or in low-mass systems, clouds have sufficiently short lifetimes that they  
are disrupted after a few free-falls times, as previously suggested (\citealt{elm00}; 
\citealt{ballhart}; \citealt{murray}).  

In this light, cloud longevity would appear to provide an unsatisfactory resolution to the issue of low observed star formation efficiencies (e.g. \citealt{kt07}).  
Instead, short-lived clouds may suggest that only a fraction of the cloud population undergoes `active star formation'.   Clouds can then be individually efficient star formers (as suggested by some models; \citealt{klessen}; \citealt{bonnell}; \citealt{VS03}; \citealt{clark05}) even while the overall efficiency in the current gas reservoir remains low.   

\subsubsection{The cycling of molecular gas from diffuse to bound objects}
Our measurements further test the notion of cloud longevity by directly contradicting several arguments invoked in its favor. 
We find that clouds more massive than 10$^5$ M$_\odot$ in M51 survive for much less time than the characteristic timescale of the molecular phase (see Figure \ref{fig:timescales}) previously upheld as a measure of the cloud lifetime.  The difference in timescales immediately suggests that a significant fraction of the molecular gas exists outside of clouds with masses 10$^5$ M$_\odot$ or more.  
Indeed, roughly half of PAWS CO emission has been determined to be in the form of an extended component, rather than in compact structures (\citealt{petyPAWS}; \citealt{colombo2014a}).  

The existence of a molecular non-cloud medium itself suggests that gas can stay molecular all the while clouds are dispersed and reassemble.  Comparing the mass in clouds with the mass in M51s non-cloud medium, we find that the conversion of the cloud component can be fast.  
We independently estimate a very short characteristic cloud timescale based on the continuity arguments in $\S$ \ref{sec:continuity}, i.e. $t_{cont}$=$M_{clouds}/M_{non-cloud}\times\tau_{non-cloud}$ where $M_{clouds}$ is the mass in cloud form, $M_{non-cloud}$ is the (molecular and atomic) mass outside clouds and $\tau_{non-cloud}$ is set to $t_{travel}$.  This timescale, which is plotted in Figure \ref{fig:measurements}, is comparable to the cloud lifetimes measured with our framework. (The average is represented by the rightmost bar in Figure \ref{fig:contprofiles}.)  
The difference may indicate that our measurements overestimate the true lifetime.   
But it more likely signifies that gas remains outside clouds, in the non-cloud medium, longer than we adopted in $\S$ \ref{sec:continuity}.  

Recall that we conservatively let $\tau_{non-cloud}$=$t_{travel}$$\sim$$t_{orb}/2$, likely underestimating the actual time it takes for the non-cloud medium to be converted into clouds by a factor of 2. 
If we take our measurements as the true lifetime, then comparing $\tau$ with $t_{cont}$ suggests that $\tau_{non-cloud}$ may be closer to $2\tau\approx2A^{-1}\lesssim2t_{travel}$, i.e. the timescale for cloud formation from the non-cloud medium is on the order of a dynamical time.  

If the reverse is true, and $t_{cont}$ is a more realistic measure of the cloud lifetime than our estimate, then eq.(\ref{eq:timescales}) implies that the inter-arm cloud population contains growth characterized by a timescale that is roughly 2$t_{cont}$=$2t_{travel}$. 
 
Interestingly, in either case, to match $t_{cont}$ and $\tau$ in Figure \ref{fig:measurements} requires that clouds leave the inter-arm cloud population quicker than they can be replenished from the non-cloud medium.  Consequently, the cloud-to-non-cloud mass ratio should vary as a function of azimuth, from one side of the galaxy to another.  As we do not observe this variation, we can conclude that, as expected, clouds form in the spiral arm and these clouds feed the inter-arm population.  To maintain the inter-arm cloud-to-non-cloud mass ratio over the course of one orbital period (from 0 to 2$\pi$), as observed, the formation timescale for clouds in the arm must be on the order of 2$\tau$.  

These considerations lead us to conclude that (large-scale) galaxy dynamics regulates the cycling of molecular material between a diffuse state and a more bound state (represented by `clouds') from which stars can form.   
Clouds emerge from the diffuse state, and their recognizability might depend not only on the molecular content of a galaxy disk but also on its dynamical character, i.e. the total (baryonic) mass and its distribution, including (non-axisymmetric) bar and spiral structures.  

\subsection{Implications for other cloud populations}
Our cloud lifetime measurements in M51 not only confirm, but also extend, the picture of short cloud lifetimes so far supported by only few existing measurements.  
By probing cloud populations under sufficiently different conditions than previously considered (spanning a larger range in radius in the disk of a spiral galaxy) we establish a basis for understanding how cloud lifetimes may be expected to vary in general. 

Here we found that everywhere the shear timescale is faster than the feedback timescale the inter-arm cloud lifetime is set by A$^{-1}$, with very little sensitivity to the rate (or amount) of star formation.  
We therefore propose that, in other circumstances, knowledge of the shear timescale should be sufficient to predict the cloud lifetime.  
Whereas feedback arguably proceeds with a universal timescale set by the $\sim$ 30 Myr lifetimes of massive OB stars, the shear timescale, which can drop below 30 Myr, determines where feedback becomes the dominant limit to the cloud lifetime.  
Elsewhere, shear itself can provide a direct constraint on the lifetime.    \\

\noindent{\it In Spiral Arms}\\
In spiral arm environments, for example, where characteristic strong streaming motions can locally reduce shear, the increase in the shear timescale suggests that spiral arm clouds can be longer-lived than their counterparts in the inter-arm.  \\

\noindent{\it In Late-type Galaxies}\\
Shear should be a similarly good predictor of cloud lifetimes throughout cloud populations, particularly in more massive disks.  Since the shear timescale depends on the shape and maximum of the circular velocity in the disk, it will likely remain shorter than the feedback timescale throughout most of all but the lowest mass galaxies.  Cloud lifetimes would be expected to increase with decreasing galaxy mass, until 
shear timescale overall exceeds the feedback timescale.  
This arguably explains why clouds in the LMC have comparably short lifetimes (20-30Myr; \citealt{kawamura}) as clouds in M51, despite the much longer shear timescale. \\

\noindent{\it In Early-type Galaxies}\\
The high shear rate in the centers of massive, early-type galaxies may increase the likelihood that some molecular gas may never have the chance to form stars, if cloud lifetimes (set by the shear timescale) become comparable to, or shorter than, the free-fall time.  This might lead to less efficient star formation and lengthened gas depletion times, such as measured in early-type galaxies with CO detections, where depletion times typically exceed those measured in normal star-forming galaxies by a factor of 2.5 \citep{davis}.  

This provides a more compelling interpretation for the lengthened $\tau_{dep}$ in such galaxies than just the shape of the rotation curve (as suggested by \citealt{davis}), considering that 
\begin{equation}
A^{-1}=\frac{t_{orb}}{\pi}\frac{1}{1-\beta}
\label{eq:sheareqn}
\end{equation}
where the rotation curve shape is parameterized as $\beta$=$d$ ln $V$/$d$ ln $R$.  While eq. (\ref{eq:sheareqn}) suggests that short $A^{-1}$ may lead to longer $\tau_{dep}$ in galaxies where $\beta$ is low, as found by \citet{davis}, it is also clear that a low $\beta$ does not always imply a short $A^{-1}$.  
This resolves the discrepancy 
noted by \citet{davis} that the $\tau_{dep}$ in the disks of nearby late-type galaxies, where $\beta$$\sim$0 but $t_{orb}$ is long, are not comparably long as in early type galaxies where $t_{orb}$ is short, but $\beta$$\sim$0 as well.  

\section{Concluding Remarks}
In this paper we present a new technique to measure the lifetimes of giant molecular clouds that tracks formation and destruction within cloud populations through cloud number statistics.  Our framework uses the travel time between spiral arms as a fiducial clock, rather than a star formation-related timescale, and yields a characteristic cloud lifetime estimate even when cloud masses are poorly determined (due to e.g. uncertainties in the CO-to-H$_2$ conversion factor).  As these ambiguities are avoided, cloud lifetimes measured with our technique can yield unique insight into the dynamical influences on clouds, using only the information that is readily accessable in current and future molecular cloud surveys.  

Our first application of the method leverages the large cloud population in the inter-arm of M51 surveyed by PAWS.  In a series of radial bins, we relate the change in cloud numbers across the inter-arm to an estimate of the average cloud lifetime at that radius.  We use our detailed knowledge of gas dynamics and massive star formation across
the PAWS field-of-view to identify two radial zones where cloud destruction is dominated by shear and star formation feedback, respectively.  
Our analysis suggests the following conclusions:\\ 

\noindent 1. GMC lifetimes in the inner disk of M51 are short, typically 20 to 30 Myr.\\

\noindent 2. Shear due to galactic differential rotation is the primary limit to cloud lifetimes when the shear timescale is shorter than the feedback timescale ($\sim$30 Myr, the average lifetime of OB stars).    \\

\noindent 3. At galactocentric radii in M51 where we expect shear to be the dominant mode of cloud disruption, the evolution in the mass and number of GMCs across the inter-arm indicates that clouds are effectively dispersed.  At larger galactocentric radii where there are strong signatures of high mass star formation, the cloud population appears to undergo transformation (i.e. an exchange of
mass between high- and low-mass clouds) rather than complete dispersal.  \\

\noindent 4. Since shear depends on the shape and maximum amplitude of the galaxy's circular velocity curve, low mass disks should contain longer-lived cloud populations than higher mass disks.  In the most massive systems and in the concentrated centers of galaxies, where short shear timescales approach the free-fall time, clouds may be so short-lived that star formation is suppressed.  \\

\noindent 5. Clouds in M51 are shorter-lived than the characteristic lifetime of molecular hydrogen, implying that molecular material can continually cycle between clouds and their (non-cloud) surroundings.   Based on our results, we suggest that conversion from a diffuse molecular phase into bound objects is regulated by large-scale galaxy dynamics.  \\

Future applications of our method to high-resolution, wide-field CO surveys of galaxies with a diverse range of Hubble types will be essential for testing these conclusions and confirming the role of short GMC lifetimes in regulating the star formation efficiency observed across a range of spatial scales.   The appropriate datasets to test our model are available with the advent of regular science operations at ALMA.

\end{document}